\documentclass[a4paper,twoside]{article}
\usepackage{richpom,graphics}
\input{epsf}
\begin{document}
%%%%%%%%%%%%%%%%%%%%%%%%%%%%%%%%%%%%%%%%%%%%%%%%%%%%%%%%%%%%%%%%%
\title{NEW EINSTEIN-MAXWELL FIELDS OF LEVI-CIVITA'S TYPE}       %
%%%%%%%%%%%%%%%%%%%%%%%%%%%%%%%%%%%%%%%%%%%%%%%%%%%%%%%%%%%%%%%%%
\author{}
\maketitle
\begin{center}
L. Richterek\footnote{E-mail: {\tt richter@risc.upol.cz}}\\\smallskip
{\slshape Department of Theoretical Physics, Palack\'y University,\\
Svobody 26, Olomouc, 771 46}\\\medskip
J. Novotn\'y\footnote{E-mail: {\tt novotny@physics.muni.cz}}\\\smallskip
{\slshape Department of General Physics, Masaryk University
}\\\medskip
J. Horsk\'y\footnote{E-mail: {\tt horsky@physics.muni.cz}}\\\smallskip
{\slshape Department of Theoretical Physics and Astrophysics, Masaryk
University,\\ Kotl\'a\v{r}sk\'a 2, Brno, 611 37, Czech Republic}
\end{center}

%
% --------------------------------------------------------------
\begin{abstract}
The method based on the Horsk\'y-Mitskievitch conjecture is applied
to the Levi-Civita vacuum metric. It is shown, that every Killing
vector is connected with a particular class of Einstein-Maxwell
fields and each of those classes is found explicitly. Some of
obtained classes are quite new. Radial geodesic motion in
constructed space-times is discussed and graphically illustrated in
the Appendix.
\end{abstract}

% -------------------------------------------------------------
\section{Introduction}\nopagebreak

There exists a lot of methods how to generate Einstein-Maxwell (EM)
fields from pure gravitational
ones~\cite{cooperst83,cooperst1,kinnersley69,KSMH,liang}. Some EM fields
were obtained by means of the Horsk\'y-Mitskievitch (HM)
conjecture~\cite{horsky89} based on the connection between the
four-potential of the electromagnetic field and the symmetries of the
spacetimes described by Killing vectors. In this paper we apply the
method outlined in~\cite{horsky89} to the vacuum Levi-Civita (LC) metric
which generally admits 3 Killing vectors and another Killing field for
two special choices of the metric parameters.

The paper is divided into the following parts: we start by resuming basic
characteristics of the LC solution, then we recapitulate the basic ideas of
the HM conjecture adopting them to the LC seed metric. Gradually, we come to
new five classes of the EM equations, each of which corresponds to one
Killing vector of the seed LC vacuum metric. Finally, we add the sixth class
which is interesting for another reason: for special values of its
parameters it reduces to the Bonnor-Melvin (BM) universe filled not by
magnetic but by electric background.

The Appendix deals in detail with radial geodesic motion in generated
spacetimes. In each case we compare the radial motion with the situation in
the seed LC solution and find out the way the radial geodesic motion
indicates the presence of singularities.

% -----------------------------------------------
\section{The LC solution}\nopagebreak
% -----------------------------------------------
\label{sec-lc}

The line element of the LC static vacuum spacetime can be written in the
Weyl form~\cite{bonnor,wang}
% -----------------------------------------------
\begin{equation}
{\rm d}s^{2}=-r^{4\sigma}{\rm d}t^{2} + r^{4\sigma (2\sigma -
1)}\left({\rm d}r^{2}+{\rm d}z^{2}\right) + C^{-2}r^{2-4\sigma}{\rm
d}\varphi^{2},
\label{eq-dslc}
\end{equation}
% -----------------------------------------------
where $\{ t,r,\varphi ,z\}$ are usual cylindrical coordinates:
$-\infty < t,z < \infty$, $r \geq 0$, $0 \leq \varphi < 2\pi$, the
hypersurfaces $\varphi=0,\ \varphi=2\pi$ are identified. The
expression (\ref{eq-dslc}) contains two arbitrary constants
$\sigma,\ C$, both of them are fixed by the internal composition of
the physical source. The constant $C$ refers to the deficit angle,
and cannot be removed by scale transformations. The physical
importance of the other parameter $\sigma$ is mostly understood in
accordance with the Newtonian analogy of the LC solution -
the gravitational field of an infinite uniform line-mass (``infinite
wire'') with the linear mass density $\sigma$~\cite{bonnor,wang}.

All orthonormal bases employed in the further calculations below
are always chosen as a generalization of the set
% -----------------------------------------------
\begin{equation}
\def\arraystretch{1.5}
\begin{array}{@{}*{4}{l}}
\bsmomega^{(0)}=r^{2\sigma}\vec{dt}, &\quad &
\bsmomega^{(1)}=r^{2\sigma (2\sigma-1)}\vec{dr},\\
\bsmomega^{(2)}=C^{-1}r^{1-2\sigma}\vec{d}\bvphi, &\quad &
\bsmomega^{(3)}=r^{2\sigma (2\sigma-1)}\vec{dz},
\end{array}
\label{eq-baselc}
\end{equation}
% -----------------------------------------------
probably the simplest tetrad one can use for the LC solution.
The Kretschmann scalar
% -----------------------------------------------
\begin{equation}
{\cal R} = R_{(\alpha)(\beta)(\mu)(\nu)}R^{(\alpha)(\beta)(\mu)(\nu)}=
64\sigma^{2}\left(4\sigma^{2}-2\sigma+1\right)
\left( 2\sigma -1\right)^{2}r^{-16\sigma^{2}+8\sigma-4},
\label{eq-Rlc}
\end{equation}
% -----------------------------------------------
where $R_{(\alpha)(\beta)(\mu)(\nu)}$ are the components of the Riemann
tensor in a chosen orthonormal basis, is infinite at $r=0$ for all $\sigma,\
C$ excluding $\sigma=0$ and $\sigma =\frac{1}{2}$ when the
spacetime is flat (see below). Thus metric (\ref{eq-dslc}) has a singularity
along the $z$-axis $r=0$ that is preferably interpreted as the infinite line
source. There is evidently no horizon, the spacetime is asymptotically
flat in the radial direction for $\sigma \neq 0,1/2$.

The analytic form of non-zero Weyl scalars
% -----------------------------------------------
\refstepcounter{equation}
$$
\eqalignno{
\Psi_{0} = \Psi_{4} & = \left( 2\sigma -1\right)\left( 2\sigma +1\right)
\sigma r^{-8\sigma^{2} + 4\sigma - 2},\label{eq-lcpsi0}
     & (\theequation.{\rm a})\cr
\Psi_{2} & = \left( 2\sigma -1\right)^{2}
\sigma r^{-8\sigma^{2} + 4\sigma - 2}\label{eq-lcpsi2}
     & (\theequation.{\rm b})\cr}
$$\label{eq-weylsclc}%
% -----------------------------------------------
leads to the conclusion that the LC metric (\ref{eq-dslc})
belongs generally to the Petrov type $I$ with the exception of
algebraically special cases belonging either to the Petrov
type $0$ or to the Petrov type $D$:
% -----------------------------------------------
% \begin{table}[h]
\begin{center}
% \caption{
%    Petrov types for special values of the parameter
%    $\sigma$ (LC solution). 
%    }
% \label{tab-lc}
\renewcommand{\arraystretch}{2}
\begin{tabular}{|c|c|c|}
\hline
$\sigma$ & Petrov type & Weyl Scalars\\
\hline
0, $\displaystyle {1\over 2}$ & 0 & all zero\\
$-\displaystyle {1\over 2}$ & $D$ & $\Psi_{2} = -\displaystyle{ 2\over
r^{6}}$\\
1 & $D$ & $\Psi_{0} = \Psi_{4} = \displaystyle{ 3\over r^{6}},\
\Psi_{2} = \displaystyle{ 1\over r^{6}}$\\
$\displaystyle{1\over 4}$ & $D$ &
$\Psi_{0} = \Psi_{4} = -\displaystyle{ 3\over 16}r^{-3/2},\
\Psi_{2} = \displaystyle{ 1\over 16}r^{-3/2}$\\[1.5ex]
\hline
\end{tabular}
\end{center}
% \end{table}
% -----------------------------------------------

As any static cylindrically symmetric solution, the metric (\ref{eq-dslc})
admits three Killing vectors:
% -----------------------------------------------
\refstepcounter{equation}
$$
\eqalignno{
\bxi_{z} &= r^{4\sigma(2\sigma-1)}\vec{dz} =
     r^{2\sigma (2\sigma-1)}\bsmomega^{(3)}
     \Longleftrightarrow \partial_{z},
     & (\theequation.{\rm a})\cr
\bxi_{\varphi} &= C^{-2}r^{2-4\sigma}\vec{d}\bvphi =
     C^{-1}r^{1-2\sigma}\bsmomega^{(2)}
     \Longleftrightarrow \partial_{\varphi},
     & (\theequation.{\rm b})\cr
\bxi_{t} &= r^{4\sigma}\vec{dt} = r^{2\sigma}\bsmomega^{(0)}
     \Longleftrightarrow \partial_{t};
     & (\theequation.{\rm c})\cr}
$$\label{eq-killc}%
% -----------------------------------------------
these Killing vectors determine the integrals of motion for geodesic
trajectories. In the case $\sigma=\frac{1}{4}$,
resp.~$\sigma=-\,\frac{1}{2}$ the LC solution has another Killing vector
$\bxi_{1/4} = -\varphi r\vec{dt} +tr\vec{d}\bvphi$, resp.~$ \bxi_{-1/2} = -z
r^{4}\vec{d}\bvphi +\varphi r^{4}\vec{dz}$. The former corresponds to a
Lorentz boost in $t-\varphi$ plane, the latter to a rotation in $\varphi-z$
plane. All mentioned Killing vectors can generate EM fields as is shown
below.

Although the value of $\sigma$ is not bound by any mathematical relation, the
physical interpretation sets strict limits. The LC solution is
traditionally identified with an infinite line source only if
$0<\sigma<\frac{1}{4}$ (note: the value $\sigma =1$ represents
$10^{28}\,\mbox{g}\cdot\mbox{cm}^{-1}$ \cite{bonnor}). Negative values
$\sigma < 0 $ lead to the negative linear density of the Newtonian analogy
and thus violate the energy conditions. Recently, it was shown
(see \cite{wang}) that even for $0\leq\sigma\leq 1 $ the LC solution
(\ref{eq-dslc}) can be produced by realistic cylindrical sources, namely
cylindrical shells of an anisotropic fluid. However, in cases
$\sigma=0,\frac{1}{2}$ when the metric is flat, the parameter $\sigma$
cannot be interpreted as a linear density at all. 

Difficulties in interpreting the LC solution were illustrated by
Bonnor~\cite{bonnor}. In case $\sigma=-1/2, C=1$ the metric (\ref{eq-dslc})
can be transformed either into Taub's plane symmetric solution, or into the
Robinson-Trautman solution, or into the solution describing the gravitational
field of a semi-infinite line-mass. Each of this possibilities suggests
a different physical interpretation. More information about the LC metric,
especially about the character of the source, can be found in
\cite{bonnor,wang} and in the references cited therein.

% -----------------------------------------------
\section{The HM conjecture and its application}\nopagebreak
% -----------------------------------------------
\label{sec-hmc}

The HM conjecture proposed in~\cite{horsky89} outlines an
efficient and fruitful way, how to obtain solutions of EM
equations as a generalization of some already known vacuum seed metrics. Its
mathematical background is based on the striking analogy between equations
satisfied by Killing vectors $\bxi$ in vacuum spacetimes
% -----------------------------------------------
\[
\vec{*d}\vec{*d}\bxi = 0,
\]
% -----------------------------------------------
and vacuum (sourceless) Maxwell equations for a testing electromagnetic
four-potential
% -----------------------------------------------
\[
\vec{*d}\vec{*dA} = 0.
\]
% -----------------------------------------------
This connection is well known for a long time and is pointed out in
many textbooks (see e.g.~\cite{lightman} p.~66,~326).

This suggestive coincidence inspired Horsk\'y and Mitskievitch
in~\cite{horsky89} to formulate the conjecture that can be expressed
in the following way (quoted verbally according to~\cite{cataldo94}):

\medskip
{\em ``The electromagnetic four-potential of a stationary self
consistent Einstein-Maxwell field is simultaneously proportional (up
to constant factor) to the Killing covector of the corresponding
vacuum spacetime when the parameter connected with the
electromagnetic field of the self-consistent problem is set equal to
zero, this parameter coinciding with the afore-mentioned constant
factor.''}
\medskip

Let $\vec{g}$ denotes the metric tensor of a vacuum seed metric, $q$
parameter characterizing the strength of the electromagnetic field
(mentioned in the quotation above) and $\tilde{\vec{g}}=\tilde{\vec{g}}(q)$
the metric tensor of an EM field, representing in fact
one-parameter class of solutions. According to the conditions of the HM
conjecture
% -----------------------------------------------
\[
\lim_{q\rightarrow 0}\tilde{\vec{g}} = \vec{g};
\]
% -----------------------------------------------
i.e. in the case of null electromagnetic field $q=0$ one comes back to the
original seed vacuum metric. The above quoted formulation of the conjecture
was further generalized by Cataldo, Kunaradtya and Mitskievitch
\cite{cataldo94} so that the four-potential $\vec{A}$ need not be inevitably
multiplied only by a constant factor, but also by a suitable scalar function
$\cal{F}$. The function is evidently not arbitrary; it must satisfy
sourceless Maxwell equations
% -----------------------------------------------
\[ \
\vec{*d*d}\left( \cal{F}\bxi\right) = 0
\]
% -----------------------------------------------
with respect to $\tilde{\vec{g}}$ (see \cite{cataldo94}). Here the
constant parameter $q$ must be involved in the analytical expression
of the function $\cal{F}$. From the point of this generalization it
is not necessary to demand that $\bxi$, a Killing vector with
respect to $\vec{g}$, must be also a Killing vector with respect to
$\tilde{\vec{g}}$ as used to be argued (see e.g. \cite{novotny92}).
This generalization leads to more vague connection between the
Killing vectors and four-potentials. At the same time, however, it
enables to find a wealth of situations in which the presumptions
of this generalized HM conjecture are fulfilled.

One can find many particular examples in which the HM conjecture works, for
example, EM fields listed in ~\cite{horsky89, horsky93, horsky94}, the
C-metric with electric field \cite{cornish1, cornish2}, most electro-vacuum
spacetimes of Petrov type $D$ studied in~\cite{KSMH} (see p.~137,
Eq.~(11.61) (Datta), p.~158-159 (McVittie, Patnaik), p.~158
(Reissner-Nordstrom) p.~297 (Kowalczynski-Plebanski)) or even quite
complicated metric studied by Kramer and Perj\'es~\cite{kramer96}. It was
also employed in generating new electromagnetic spacetimes from vacuum
metrics~\cite{cataldo94, novotny92, stefanik}. Its applicability and
possible limitations are still under systematic investigation. Although the
original formulation quoted above restricts itself only to vacuum seed
spacetimes $\vec{g}$ and static or stationary EM fields $\tilde{\vec{g}}$,
nowadays these conditions seem to become redundant. The possibility to
construct new solutions also from non-vacuum seed metrics is supposed
immediately in~\cite{horsky89}. In section \ref{sec-elms4} a non-stationary
solution will be obtained through the procedure of the generalized
conjecture.

Probably the most important advantage of the HM conjecture is the
opportunity to choose the character of the electromagnetic field we would
like to obtain. It is obviously determined by the vector potential and thus
by the geometrical substance of the Killing vector one uses for the
generation of $\tilde{\vec{g}}$. If the seed metric $\vec{g}$ admits more
than one Killing vector, there is usually possible to construct more EM
fields, each of them corresponding to a different Killing vector. The
situation is most lucid when the seed metric $\vec{g}$ is static. In that
case a straightforward calculation leads to the conclusion that rotational,
as well as space-like translational Killing vectors, give magnetic EM fields,
while timelike translational and the boost Killing vectors lead to the
electric fields (in full analogy with the Minkowski spacetime). For the
rotational Killing vector $\partial_{\varphi}$ in common cylindrical
coordinates the correspondence with longitudinal magnetic field was
demonstrated by Wald (see e.g. \cite{lightman}, p.~66,~326). Let us remind
that the electric or magnetic character of any obtained EM field determines
the sign of the electromagnetic invariant
% -----------------------------------------------
\begin{equation}
F_{(\mu)(\nu)}F^{(\mu)(\nu)} = 2\left( B^{2} - E^{2}\right) ,
\label{invariant}
\end{equation}
% -----------------------------------------------
where $\vec{F}$ is an antisymmetric tensor of the electromagnetic field
related to the components of electric and magnetic field strengths $\vec{E}$
and $\vec{B}$ in a standard way (see e.g.~\cite{lightman},~p.~23,
\cite{MTW},~p.~74). If $F_{(\mu)(\nu)}F^{(\mu)(\nu)} > 0$, the field is of
magnetic type, if $F_{(\mu)(\nu)}F^{(\mu)(\nu)} < 0$ then it is of electric
type.

% The most difficult and still unresolved problem related to the HM
% conjecture is the absence of an explicit step-by-step algorithm for
% constructing new EM fields; in excessive generality Einstein
% and sourceless Maxwell equations lead to rather complicated unsolvable
% formulae. The key question is, what analytic form of $\tilde{\vec{g}}$ one
% should assume to get any reasonable results and solvable equations.
% Therefore, at nowadays stage finding a new EM field by means
% of the HM conjecture inevitably requires some intuition and the result is
% usually obtained after several trials and impasses.
%
Spacetimes treated below in this paper were found through the HM conjecture
from the LC seed metric. The procedure leads to six new solutions containing
different types of the electromagnetic field. For the sake of efficiency all
tensor components were expressed in tetrad formalism. Retrospectively, it
was possible to trace back the analogous steps in calculations and resume
them in the following scheme:
% -----------------------------------------------
\begin{itemize}
\item[a)] According to (\ref{eq-baselc}) and (\ref{eq-killc}) Killing
vectors of $\bxi_{z},\ \bxi_{\varphi},\ \bxi_{t}$ are collinear with one of
the basis vectors (\ref{eq-baselc}). It is possible to choose the vector
potential $\vec{A}$ in such a way that its {\em tetrad} components with
respect $\tilde{\vec{g}}$ coincide - up to a constant factor $q$
characterizing the strength of the electromagnetic field - with the
components of the Killing vectors with respect to $\vec{g}$, i.e. with
respect to (\ref{eq-baselc}). In the case of $\bxi_{1/4}$ and $\bxi_{-1/2}$
(see sections
\ref{sec-elms4},~\ref{sec-elms2}) the tetrad vectors were chosen in
such a way that one of the basis vectors coincides either with $\bxi_{1/4}$
or with $\bxi_{-1/2}$. Then, for simplicity's sake a suitable coordinate
transformation was performed.

\item[b)] The question is, in what way we should modify
$\tilde{\vec{g}}$ in relation to $\vec{g}$ or (in tetrad formalism), in what
way the orthonormal basis used for $\tilde{\vec{g}}$ should differ from that
one used for $\vec{g}$. In all cases below one basis vector from
(\ref{eq-baselc}), namely that one corresponding to the vector potential, is
divided, and the others multiplied by the same function $f(t,r,\varphi,z)$.
The conditions of HM conjecture are fulfilled if
% -----------------------------------------------
\begin{equation}
f(t,r,\varphi,z) = 1+c_{1}f_{1}(t,r,\varphi,z),
\label{eq-genf}
\end{equation}
% -----------------------------------------------
and the function $f_{1}(t,r,\varphi,z)$ must be a solution of the
differential equation
% -----------------------------------------------
\begin{equation}
G=G^{(\mu)}_{\ \ \ (\mu)}= -R =0
\label{eq-treinst}
\end{equation}
% -----------------------------------------------
arising from the well known fact that for a pure electromagnetic
spacetime the Einstein tensor is traceless. Thus all obtained solutions
have zero scalar curvature $R$. We shall see that the analytic form
of $f_{1}$ coincides with the basis vector (\ref{eq-baselc}) collinear
with the vector potential. The constant $c_{1}$ in (\ref{eq-genf})
must naturally involve the parameter $q$ as the limit
% -----------------------------------------------
\[
 \lim_{c_{1}\rightarrow 0} f(t,r,\varphi,z) = 1
\]
% -----------------------------------------------
for any regular $f_{1}$ gives original seed metric.
\item[c)] Completing the steps a) and b) we can ensure the validity of
sourceless Maxwell equations. Substituting $\tilde{\vec{g}}$ into the
Einstein equations we are able to fit the constant $c_{1}$ against $q$.
\end{itemize}
% -----------------------------------------------

Although the simplicity of the above outlined algorithm is obviously caused
by relatively high degree of symmetry characteristic for the LC metric,
this scheme might contribute to the discussion about the application of the
conjecture and the character of relation between Killing vectors and
electromagnetic field. The following sections are devoted to the
particular applications of the outlined scheme.

% -----------------------------------------------
\section{The LC solution with azimuthal magnetic field}\nopagebreak
% -----------------------------------------------
\label{sec-mgwaz}

Let us start with the Killing vector $\bxi_{z}$.  The simplest possible
choice of $f$ in Eq.~(\ref{eq-genf}) is $f=f(r)$. Therefore, in spirit of
the above discussed scheme the modified tetrad takes the the form
% -----------------------------------------------
\begin{equation}
\def\arraystretch{1.5}
\begin{array}{@{}*{4}{l}}
\displaystyle\bsmomega^{(0)}=f(r)r^{2\sigma}\vec{dt}, &\quad &
\displaystyle\bsmomega^{(1)}=f(r)r^{2\sigma (2\sigma-1)}\vec{dr},\\
\displaystyle\bsmomega^{(2)}=f(r)C^{-1}r^{1-2\sigma}\vec{d}\bvphi,&\quad &
\displaystyle\bsmomega^{(3)}=\frac{r^{2\sigma (2\sigma-1)}}{f(r)}\vec{dz}
\end{array}
\label{eq-basemgph}
\end{equation}
% -----------------------------------------------
and the corresponding line element reads as
% -----------------------------------------------
\begin{equation}
{\rm d}s^{2}=-f(r)^{2}r^{4\sigma}{\rm d}t^{2} +
f(r)^{2}r^{4\sigma (2\sigma - 1)}{\rm d}r^{2} +
f(r)^{2}C^{-2}r^{2-4\sigma}{\rm d}\varphi^{2}
+ \frac{r^{4\sigma (2\sigma - 1)}}{f(r)^{2}}{\rm d}z^{2}.
\label{eq-dsmgph}
\end{equation}
% -----------------------------------------------
For the four-potential we have
% -----------------------------------------------
\begin{equation}
\vec{A} = q\frac{r^{4\sigma (2\sigma -1)}}{f(r)}\,\vec{dz} =
qr^{2\sigma (2\sigma -1 )}\bsmomega^{(3)}.
\label{eq-vpmgph}
\end{equation}
% -----------------------------------------------
The condition of traceless Einstein tensor (\ref{eq-treinst})
determines the function $f$
% -----------------------------------------------
\begin{equation}
f(r) = 1 + c_{1}r^{4\sigma (2\sigma - 1)}
\label{eq-fcemgph}
\end{equation}
% -----------------------------------------------
and from the Einstein equations one obtains
% -----------------------------------------------
\[
c_{1}=q^{2}.
\]
% -----------------------------------------------
% Consequently, non-zero tetrad components of Einstein tensor are
% -----------------------------------------------
% \begin{equation}
% G_{(0)(0)}=G_{(1)(1)}=-\,G_{(2)(2)}=G_{(3)(3)} =
% 16\sigma^{2}q^{2}\frac{\left(2\sigma - 1\right)^{2}}
% {r^{2}f(r)^{4}}.
% \label{eq-einstmgph}
% \end{equation}
% -----------------------------------------------
The electromagnetic field is of magnetic type, which can be
demonstrated by the electromagnetic invariant (\ref{invariant})
% -----------------------------------------------
\begin{equation}
F_{(\mu )(\nu )}F^{(\mu )(\nu )} =
32\,\frac{q^{2}\sigma^{2}\left( 2\sigma-1\right)^{2}}
{r^{2}f(r)^{2}} \geq 0
\label{eq-invmgph}
\end{equation}
% -----------------------------------------------
or directly by the electromagnetic field tensor
% -----------------------------------------------
\begin{equation}
F^{(1)(3)} = - B^{(2)} =
\,\frac{4q\sigma (2\sigma -1)}{rf(r)^{2}}.
\label{eq-elmgmgph}
\end{equation}
% -----------------------------------------------
The only non-zero component of the magnetic field strength $\vec{B}$ is
the azimuthal one. In accordance with the accepted physical interpretation
of the LC solution (at least for some values of $\sigma$) the line
element (\ref{eq-dsmgph}) describes an EM field of an
infinite line source with an azimuthal magnetic field, in other words, the
gravitational field of an infinite line source with an electric
current.

The metric (\ref{eq-dsmgph}) has a one-dimensional singularity along
the $z$-axes $r=0$ where the Kretschmann scalar
% -----------------------------------------------
\begin{equation}
\def\arraystretch{1.5}
\begin{array}{@{}*{4}{l}}
{\cal R} &=&\displaystyle
\frac{64\sigma^{2}\left(2\sigma
-1\right)^{2}}{f(r)^{8}r^{4(4\sigma^{2}-2\sigma+1)}}
\left. \rule{0pt}{3ex}\right[g_{1}(r)^{4}
\left(4\sigma-1\right)^{2}\left(12\sigma^{2}-6\sigma+1\right)-
\\
&&\displaystyle -\,{}12 g_{1}(r)^{3}\sigma
\left( 2\sigma-1\right)\left(4\sigma-1\right)^{2} +
\\
&&\displaystyle +\,{}2g_{1}(r)^{2}
\left(160\sigma^{4}-160\sigma^{3}+24\sigma^{2}+8\sigma-1\right) +\\
&&\displaystyle+\,{}12g_{1}(r)\sigma 
\left(2\sigma-1\right) +
4\sigma^{2} -2\sigma + 1
\left. \rule{0pt}{3ex}\right].
\end{array}
\label{eq-Rmgph}
\end{equation}
% -----------------------------------------------
becomes infinite; $g_{1}(r) = q^{2}r^{4\sigma (2\sigma -1)}$.
Obviously, the only exceptions are the cases $\sigma=0$,
$\sigma=\frac{1}{2}$ for which the seed LC metric is flat.
Then also the metric (\ref{eq-dsmgph}) becomes flat and
does not include any electromagnetic field.

Eventually, the analytic expressions for the Weyl scalars have the form
% -----------------------------------------------
\refstepcounter{equation}
$$
\eqalignno{
\Psi_{0} = \Psi_{4} = & - \frac{\left(2\sigma
-1\right)\sigma}{r^{2}f(r)^{5}g_{1}(r)}\left(q^{4}g_{1}(r)^{2}-1\right)
\left[q^{2}g_{1}(r)\left(24\sigma^{2}-10\sigma+1\right) + \right.& \cr
& +\,{}\left.\left( 2\sigma +1\right)\right],
& (\theequation.{\rm a})\cr
\Psi_{2} = & \,{}\frac{\left(2\sigma
-1\right)^{2}\sigma}{r^{2}f(r)^{7}g_{1}(r)}
\left.\rule{0pt}{3ex}\right[
q^{10}g_{1}(r)^{5}(4\sigma - 1) + q^{8}g_{1}(r)^{4}(8\sigma -3) - & \cr 
& -\,{}2q^{6}g_{1}(r)^{3} -
2q^{4}g_{1}(r)^{2}(4\sigma -1) - q^{2}g_{1}(r)(4\sigma -3) + 1
\left.\rule{0pt}{3ex}\right]. & (\theequation.{\rm b})\cr} 
$$\label{eq-weylscmgph}%
% -----------------------------------------------
Thus the spacetime (\ref{eq-dsmgph}) is generally Petrov type $I$, special
cases being:
% -----------------------------------------------
% \begin{table}[h]
% \caption{Petrov types for special values of the parameter
% $\sigma$ (The LC solution with azimuthal magnetic field).}
% \label{tab-mgph}
\begin{center}
\renewcommand{\arraystretch}{2}
\begin{tabular}{|c|c|c|}
\hline
$\sigma$ & Petrov type & Weyl Scalars\\
\hline
0, $\displaystyle{1\over 2}$ & 0 & all zero\\
$\displaystyle {1\over 4}$ & $D$ &
$\Psi_{0} = \Psi_{4} = -3\Psi_{2} =
\displaystyle{ \frac{3}{16} \frac{\left( q^{2}-\sqrt{r}\right)}
{\left( q^{2}+\sqrt{r}\right)^{4}}}$
\\[1.5ex] \hline
\end{tabular}%
\end{center}%
% \end{table}
% -----------------------------------------------

It should be noted that though (\ref{eq-dsmgph}) reminds of the
general static cylindrically symmetric solution with azimuthal
magnetic field \cite{KSMH},~\S 20.2,~Eq.~20.9a, it does {\em not}
belong to this class of spacetimes. This inevitably means the
metric 20.9a in \cite{KSMH} does not represent the most general
cylindrical symmetric EM solution with azimuthal magnetic
field. Putting $q=0$, which means no electromagnetic field, one obtains the
seed LC metric.

% -----------------------------------------------
\section{The LC solution with longitudinal magnetic field}\nopagebreak
% -----------------------------------------------
\label{sec-mgwz}

Let us take the Killing vector $\bxi_{\varphi}$ now. This requires the
tetrad
% -----------------------------------------------
\begin{equation}
\def\arraystretch{1.5}
\begin{array}{@{}*{4}{l}}
\displaystyle\bsmomega^{(0)}=f(r)r^{2\sigma}\vec{dt}, & \quad &
\displaystyle\bsmomega^{(1)}=f(r)r^{2\sigma (2\sigma-1)}\vec{dr},\\
\displaystyle\bsmomega^{(2)}=\frac{r^{1-2\sigma}}{C f(r)}\vec{d}\bvphi, 
             & \quad &
\displaystyle\bsmomega^{(3)}=f(r)r^{2\sigma (2\sigma-1)}\vec{dz}
\end{array}
\label{eq-basemgz}
\end{equation}
% -----------------------------------------------
and leads to the line element
% -----------------------------------------------
\begin{equation}
{\rm d}s^{2}=-f(r)^{2}r^{4\sigma}{\rm d}t^{2} + f(r)^{2}r^{4\sigma
(2\sigma - 1)}\left[ {\rm d}r^{2} + {\rm d}z^{2}\right] +
\frac{r^{2-4\sigma}}{f(r)^{2}C^{2}}\,{\rm d}\varphi^{2}.
\label{eq-dsmgz}
\end{equation}
% -----------------------------------------------
Killing vector $\bxi_{\varphi}$ induces the four-potential
% -----------------------------------------------
\begin{equation}
\vec{A} = \frac{qr^{2(1-2\sigma )}}{C^2f(r)}\,\vec{d}\bvphi =
qC^{-1}r^{1-2\sigma }\bsmomega^{(2)}.
\label{eq-vpmgz}
\end{equation}
% -----------------------------------------------
Following the scheme described in the section~\ref{sec-hmc} one gets
% -----------------------------------------------
\begin{equation}
f(r) = 1 + c_{1}r^{2(1 - 2\sigma )},\qquad c_{1}=\frac{q^{2}}{C^2}.  
\label{eq-fcemgz}
\end{equation}
% -----------------------------------------------
% \begin{equation}
% G_{(0)(0)}=G_{(1)(1)}=G_{(2)(2)}=-\,G_{(3)(3)} =
% \frac{4q^{2}(2\sigma -1)^{2}}{C^{2}f(r)^{4}r^{8\sigma^{2}}}.
% \label{eq-einstmgz}
% \end{equation}
% -----------------------------------------------
The electromagnetic field is again of magnetic type since
% -----------------------------------------------
\begin{equation}
F_{(\mu )(\nu )}F^{(\mu )(\nu )} =
\frac{8q^{2}\left( 2\sigma-1\right)^{2}}
{C^2f(r)^{4}r^{8\sigma^{2}}} \geq 0,
\label{eq-invmgz}
\end{equation}
% -----------------------------------------------
while the only non-zero component of magnetic field strength being the
longitudinal one in direction along the $z$-axis
% -----------------------------------------------
\begin{equation}
F^{(1)(2)} = B^{(3)} =
-\frac{2q(2\sigma -1)}{Cf(r)^{2}r^{4\sigma^{2}}}.
\label{eq-elmgmgz}
\end{equation}
% -----------------------------------------------
In accordance with the accepted physical interpretation of LC solution (at
least for some values of $\sigma$) the line element (\ref{eq-dsmgz})
describes an EM field of an infinite line source with a longitudinal
magnetic field, in other words, the gravitational field of an infinite line
source in a Bonnor-Melvin-like universe (see e.g.~\cite{KSMH}, \S~20.2,
Eq.~20.10) which is responsible for the background longitudinal magnetic
field.  The situation reminds us of the solution described by Cataldo et. all
\cite{cataldo94} called ``pencil of light in the Bonnor-Melvin Universe''.
Moreover, substituting $\sigma=0$ into (\ref{eq-dsmgz}) one easily comes to
% -----------------------------------------------
\begin{equation}
{\rm d}s^{2}= \left(1+q^2r^{2}/C^{2}\right)^{2}
\left[ -{\rm d}r^{2} + {\rm d}r^{2} + {\rm d}z^{2}\right] +
\frac{r^{2}}{\left(1+q^2r^{2}/C^{2}\right)C^{2}}\,{\rm d}\varphi^{2}.
\label{eq-dsbonmel}
\end{equation}
% -----------------------------------------------
Evidently, this metric with $q=B_{0}/2$ and $C=1$ gives the well-known
Bonnor-Melvin solution of EM equations.  Let us remind that for
$\sigma=0,\ C=1$ the seed LC metric reduces to Minkowski spacetime.
Thus, the Bonnor-Melvin universe can be obtained through the HM conjecture
straight from the Minkowski spacetime expressed in common cylindrical
coordinates. This possibility was already mentioned by Cataldo
at. all~\cite{cataldo94}.

Like the seed LC metric, the solution (\ref{eq-dsmgz}) has
one-dimensional singularity along the $z$-axis, since the
Kretschmann scalar
% -----------------------------------------------
\begin{equation}
\def\arraystretch{1.5}
\begin{array}{@{}*{4}{l}}
{\cal R} &=&\displaystyle
\frac{64\left(2\sigma -1\right)^{2}}
{f(r)^{8}r^{8\sigma (2\sigma -1)+4}}
\left. \rule{0pt}{3ex}\right[
g_{2}(r)^{4}
\left( 4\sigma^{2}-6\sigma+3 \right)\left(\sigma -1\right)^{2}+
\\
&&\displaystyle+\,{}6g_{2}(r)^{3}
\left( 2\sigma -1\right)\left(\sigma -1\right)^{2} -\\
&&\displaystyle -g_{2}(r)^{2}
\left( 8\sigma^{4}-16\sigma^{3}-12\sigma^{2}+ 20\sigma -5\right) -\\
&&-\,{}6g_{2}(r)\sigma^{2}
\left(2\sigma -1\right) + \sigma^{2}\left(4\sigma^{2}-2\sigma +1\right)
\left.\rule{0pt}{3ex}\right],
\end{array}
\label{eq-Rmgz}
\end{equation}
% -----------------------------------------------
where $g_{2}(r)= q^{2}r^{2-4\sigma}/C^{2}$, diverges at $r=0$. The
only exception is clearly the case $\sigma=\frac{1}{2}$ for which the
metric is flat and does not include any electromagnetic field.

The expressions for the Weyl scalars
% -----------------------------------------------
\refstepcounter{equation}
$$
\eqalignno{
\Psi_{0} = \Psi_{4} = & \frac{\left(2\sigma
-1\right)\left(C^{2}r^{4\sigma}-q^{2}r^{2}\right)}
{C^{4}f(r)^{4}r^{8\sigma^{2}-12\sigma+2}}
\left[r^{4\sigma}C^{2}\sigma\left(2\sigma+1\right)+\right. & \cr
& + \left. q^{2}r^{2}\left(2\sigma^{2}-5\sigma+3\right)\right],
& (\theequation.{\rm a})\cr   
\Psi_{2} = & \frac{\left(2\sigma
-1\right)^{2}\left(C^2r^{4\sigma}-q^{2}r^{2}\right)}
{C^{4}f(r)^{4}r^{8\sigma^{2}-12\sigma+2}}
\left[
C^{2}\sigma r^{4\sigma} + q^{2}r^{2}\left(\sigma -1\right) \right].
     & (\theequation.{\rm b})\cr}
$$\label{eq-weylscmgz}%
% -----------------------------------------------
again leads to the conclusion that the metric is generally Petrov type $I$
with the exception of algebraically special cases
% -----------------------------------------------
% \begin{table}[h]
% \caption{Petrov types for special values of the parameter
%    $\sigma$ (The LC solution with longitudinal magnetic field).}
% \label{tab-mgz}
\begin{center}
\renewcommand{\arraystretch}{2}
\begin{tabular}{|c|c|c|}
\hline
$\sigma$ & Petrov type & Weyl Scalars\\
\hline
0 & $D$ &
$\Psi_{0} = \Psi_{4} = 3\Psi_{2} =
\displaystyle{\frac{3q^{2}\left( q^{2}r^{2}-C^{2}\right)}
{C^{4}\left(C^{2}+q^{2}r^{2}\right)^{4}}}$\\
$\displaystyle{1\over 2}$ & 0 & all zero\\
1 & $D$ &
$\Psi_{0} = \Psi_{4} = 3\Psi_{2} =
\displaystyle{\frac{3\left( C^{2}r^{2}-q^{2}\right)}
{C^{2}\left(C^{2}r^{2}+q^{2}\right)^{4}}}$
\\[1.5ex] \hline
\end{tabular}
\end{center}
% \end{table}
% -----------------------------------------------

The metric (\ref{eq-dsmgz}) reminds of the general static cylindrically
symmetric solution with longitudinal magnetic field
\cite{KSMH},~\S 20.2,~Eq.~20.9b, but it does {\em not} belong to this
class of spacetimes (the only exception being the Bonnor-Melvin universe
(\ref{eq-dsbonmel})). This definitely means that the metric 20.9b in
\cite{KSMH} does not represent the most general cylindrical symmetric
EM field with longitudinal magnetic field. 

% -----------------------------------------------
\section{The LC solution with radial electric field}\nopagebreak
% -----------------------------------------------
\label{sec-elw}

The last from the Killing vectors (\ref{eq-killc}) is $\bxi_{t}$. The
tetrad
% -----------------------------------------------
\begin{equation}
\def\arraystretch{1.5}
\begin{array}{@{}*{4}{l}}
\displaystyle\bsmomega^{(0)}=\frac{r^{2\sigma}}{f(r)}\vec{dt}, &\quad &
\displaystyle\bsmomega^{(1)}=f(r)r^{2\sigma (2\sigma-1)}\vec{dr},\\
\displaystyle\bsmomega^{(2)}=f(r)C^{-1}r^{1-2\sigma}\vec{d}\bvphi, &\quad &
\displaystyle\bsmomega^{(3)}=f(r)r^{2\sigma (2\sigma-1)}\vec{dz}
\end{array}
\label{eq-baseelw}
\end{equation}
% -----------------------------------------------
determines the metric
% -----------------------------------------------
\begin{equation}
{\rm d}s^{2}=-\frac{r^{4\sigma}}{f(r)^{2}}{\rm d}t^{2} +
f(r)^{2}r^{4\sigma (2\sigma - 1)}\left[{\rm d}r^{2} + {\rm
d}z^{2}\right] + f(r)^{2}C^{-2}r^{2-4\sigma}{\rm d}\varphi^{2},
\label{eq-dselw}
\end{equation}
% -----------------------------------------------
where
% -----------------------------------------------
\begin{equation}
f(r) = 1 + c_{1}r^{4\sigma}, 
\label{eq-fceelw}
\end{equation}
% -----------------------------------------------
and
% -----------------------------------------------
\[
c_{1}=-q^{2}.
\]
% -----------------------------------------------
% The non-zero tetrad components of Einstein tensor then read
% -----------------------------------------------
% \begin{equation}
% G_{(0)(0)}=-G_{(1)(1)}=G_{(2)(2)}=G_{(3)(3)} =
% \frac{16q^{2}\sigma^{2}r^{-8\sigma^{2}+8\sigma-2}}{f(r)^{4}}.
% \label{eq-einstelw}
% \end{equation}
% -----------------------------------------------
The vector potential
% -----------------------------------------------
\begin{equation}
\vec{A} = -\frac{qr^{4\sigma}}{f(r)}\vec{dt} = -qr^{2\sigma}\bsmomega^{(0)}
\label{eq-vpelw}
\end{equation}
% -----------------------------------------------
sets the field of electric type
% -----------------------------------------------
\begin{equation}
F_{(\mu )(\nu )}F^{(\mu )(\nu )} =
-32\,\frac{q^{2}\sigma^{2} r^{-8\sigma^{2}+8\sigma - 2}}
{f(r)^{4}} \leq 0
\label{eq-invelw}
\end{equation}
% -----------------------------------------------
with non-zero radial component of electric field strength
% -----------------------------------------------
\begin{equation}
F^{(0)(1)} = E^{(1)} =
-\,\frac{4q\sigma r^{-4\sigma^{2}+4\sigma-1}}{f(r)^{2}}.
\label{eq-elmgelw}
\end{equation}
% -----------------------------------------------
One can conclude the metric (\ref{eq-dselw}) describes
EM field of a charged infinite line source.

Unlike the seed metric (\ref{eq-dslc}) and the solutions (\ref{eq-dsmgph}),
(\ref{eq-dsmgz}), the spacetime (\ref{eq-dselw}) contains not only one
dimensional singularity along the $z$-axes but also a singularity at a
radial distance $r_{s}$ for which
% -----------------------------------------------
\[
f(r_{s}) = 1-q^{2}r_{s}^{4\sigma} = 0.
\]
% -----------------------------------------------
The Kretschmann scalar
% -----------------------------------------------
\begin{eqnarray}
{\cal R} &=&
\frac{64\sigma^{2}}{f(r)^{8}r^{16\sigma^{2}-8\sigma+4}}
\left. \rule{0pt}{3ex}\right[
g_{3}(r)^{4}
\left( 4\sigma^{2}+2\sigma+1\right)\left(2\sigma +1\right)^{2}+
\nonumber\\
&& +\, 12g_{3}(r)^{3}\sigma
\left( 2\sigma+1\right)^{2} -
2g_{3}(r)^{2}
\left(16\sigma^{4}-48\sigma^{2}+1\right) -\\
&&-\,12 g_{3}(r)\sigma
\left(2\sigma+1\right)^{2} +
\left(2\sigma -1\right)^{2}\left(4\sigma^{2}-2\sigma +1\right)
\left.\rule{0pt}{3ex}\right],\nonumber
\label{eq-Relw}
\end{eqnarray}
% -----------------------------------------------
where $g_{3}(r)=q^{2}r^{4\sigma}$, proves that this surface represents
physical singularity that cannot be removed by any coordinate
transformation.

The solution (\ref{eq-dselw}) is generally Petrov type $I$ with Weyl
scalars
% -----------------------------------------------
\refstepcounter{equation}
$$
\eqalignno{
\Psi_{0} = \Psi_{4} = &\left( 2\sigma -1\right)\left( 2\sigma +1\right)
\sigma r^{-8\sigma^{2} + 4\sigma - 2}
\frac{1+q^{2}r^{4\sigma}}{f(r)^{3}},
     & (\theequation.{\rm a})\cr
\Psi_{2} = &-\,\frac{\sigma r^{-8\sigma^{2} + 4\sigma -
2}}{f(r)^{4}}
\left[q^{2}r^{4\sigma}(2\sigma + 1)^{2} -(2\sigma - 1)^{2}-
8q^{2}\sigma r^{4\sigma }\right];
     & (\theequation.{\rm b})\cr}
$$\label{eq-weylscelw}%
% -----------------------------------------------
the only exceptions belonging to other Petrov classes are
% -----------------------------------------------
% \begin{table}[h]
% \caption{Petrov types for special values of the parameter
%    $\sigma$ (The LC solution with radial electric field).}
% \label{tab-elw}
\begin{center}
\renewcommand{\arraystretch}{2}
\begin{tabular}{|c|c|c|}
\hline
$\sigma$ & Petrov type & Weyl Scalars\\
\hline
0 & 0 & all zero\\
$\displaystyle {1\over 2}$ & $D$ &
$\Psi_{2} = -\,2\,\displaystyle{ q^{2}\left(1+q^{2}r^{2}\right)\over
{\left( 1-q^{2}r^{2}\right)^{4} }}$\\
$-\,\displaystyle {1\over 2}$ & $D$ &
$\Psi_{2} = -\,2\,\displaystyle{\left( q^{2}+r^{2}\right)\over
{\left( q^{2}-r^{2}\right)^{4} }}$
\\[1.5ex]
\hline
\end{tabular}
\end{center}
% \end{table}
% -----------------------------------------------
The metric (\ref{eq-dselw}) represents a special case of the general
cylindrically symmetric solutions with radial electric field
\cite{KSMH},~\S 20.2,~Eq.~20.9c. Nevertheless, the approach within the
framework of the HM conjecture provides a promising possibility for its
physical interpretation and for understanding the nature of sources (at
least for some values of $\sigma$).

% -----------------------------------------------
\section{The magnetovacuum solution for $\sigma=-1/2$}\nopagebreak
% -----------------------------------------------
\label{sec-elms2}

Before proceeding on the Killing vector $\bxi_{-1/2}$ several explanatory
notes should be added. The vector $\bxi_{-1/2}$ obviously generates rotation
in the $\varphi-z$ plane, so it is geometrically equivalent to
$\bxi_{\varphi}$ and should lead to magnetic EM field. The cylindrical
coordinates used so far are very convenient for the expression of
$\bxi_{\varphi}$. To follow the scheme outlined in section~\ref{sec-hmc} one
should preferably start with the coordinate transformation
% -----------------------------------------------
$$
\varphi=CX\cos Y,\qquad z=X\sin Y,
$$
% -----------------------------------------------
which turn the LC metric (\ref{eq-dslc}) for
$\sigma=-\frac{1}{2}$ into the form
% -----------------------------------------------
$$
{\rm d}s^{2}=-\,\frac{{\rm d}t^{2}}{r^2} + r^{4}{\rm d}r^{2}+r^{4}{\rm
d}X^{2}+r^{4}X^{2}{\rm d}Y^{2}
$$
% -----------------------------------------------
and the Killing vector
% -----------------------------------------------
\[
\bxi_{-1/2} = \partial_{Y} = r^{4}X^{2}\vec{dY}.
\]
% -----------------------------------------------
Now in analogy with all preceding cases let us take the basis
% -----------------------------------------------
\begin{equation}
\def\arraystretch{1.5}
\begin{array}{@{}*{4}{l}}
\displaystyle\bsmomega^{(0)}=\frac{F(X,r)}{r}\vec{dt}, &\quad &
\displaystyle\bsmomega^{(1)}=F(X,r)r^{2}\vec{dr},\\
\displaystyle\bsmomega^{(2)}=F(X,r)r^{2}\vec{dX}, &\quad &
\displaystyle\bsmomega^{(3)}=\frac{r^{2}X}{F(X,r)}\vec{dY}
\end{array}
\label{eq-baseelms2a}
\end{equation}
% -----------------------------------------------
inducing a metric
% -----------------------------------------------
$$
{\rm d}s^{2}=-\,\frac{F(X,r)^2}{r^{2}}{\rm d}t^{2} + F(X,r)^{2}r^{4}{\rm
d}r^{2} +F(X,r)^{2}r^{4}{\rm d}X^{2}+\frac{r^{4}X^{2}}{F(X,r)^{2}}{\rm
d}Y^{2}
$$
% -----------------------------------------------
and set the four-potential
% -----------------------------------------------
\begin{equation}
\vec{A} = q\frac{r^{4}X^{2}}{F(X,r)}\vec{dY}.
\label{eq-vpelms2a}
\end{equation}
% -----------------------------------------------
The solution of Einstein and sourceless Maxwell equations then provides
% -----------------------------------------------
% \begin{equation}
$$ F(X,r) = 1 + q^{2}X^{2}r^{4}.$$
% \label{eq-fceelms2a}
% \end{equation}
% -----------------------------------------------
Although the coordinates $(t,r,X,Y)$ are extremely suitable for calculation of
tensor components and solving EM equations, the cylindrical
coordinates $(t,r,\varphi,z)$ fit better the aim of physical interpretation,
namely because of the evident relation to the LC seed metric.
Therefore, let us consequently transform all above computed objects back into
the cylindrical coordinates. We obtain
% -----------------------------------------------
\begin{equation}
\vec{A} = q\frac{r^{4}}{Cf(r,\varphi,z)}
\left(-z{\vec d}\bvphi+\varphi{\vec dz}\right)=
qr^{2}\sqrt{\varphi^{2}/C^2+z^{2}}\bsmomega^{(3)},
\label{eq-vpelms2}
\end{equation}
% -----------------------------------------------
% -----------------------------------------------
\begin{equation}
\def\arraystretch{1.5}
\begin{array}{@{}*{4}{l}}
\displaystyle\bsmomega^{(0)}=\frac{f(r,\varphi,z)}{r}\vec{dt},\quad 
\displaystyle\bsmomega^{(1)}=f(r,\varphi,z)r^{2}\vec{dr},\\
\displaystyle\bsmomega^{(2)}=\frac{r^{2}f(r,\varphi,z)}{\sqrt{\varphi^{2}/C^2
   +z^{2}}}\left(\frac{\varphi}{C^2}{\vec d}\bvphi+z{\vec dz}\right),\\
\displaystyle\bsmomega^{(3)}=\frac{r^{2}}{\sqrt{\varphi^{2}/C^2+
   z^{2}}f(r,\varphi,z)}\left(-\frac{z}{C}{\vec d}\bvphi+
                              \frac{\varphi}{C}{\vec dz}\right);
\end{array}
\label{eq-baseelms2}
\end{equation}
% -----------------------------------------------
and
% -----------------------------------------------
\begin{equation}
\def\arraystretch{1.5}
\begin{array}{@{}*{4}{l}}
 {\rm d}s^{2} & = &-\displaystyle{\frac{f(r,\varphi,z)^{2}}{r^{2}}}{\rm d}t^{2}
+ f(r,\varphi,z)^{2}r^{4}{\rm d}r^{2} +\\
&&\displaystyle +\,{}\frac{r^{4}}{\varphi^{2}/C^2+z^{2}}
\left[f(r,\varphi,z)^{2}\left(\frac{\varphi}{C^2}{\rm d}\varphi+
                              z{\rm d}z\right)^{2}\right.+\\
&&\displaystyle\left.+\frac{1}{f(r,\varphi,z)^{2}}
      \left(-\frac{z}{C}{\rm d}\varphi+\frac{\varphi}{C^2}{\rm d}z\right)^{2}
\right],
\end{array}
\label{eq-dselms2}
\end{equation}
% -----------------------------------------------
where
% -----------------------------------------------
\begin{equation}
f(r,\varphi,z) = 1 + q^{2}\left(\varphi^{2}/C^2+z^{2}\right)r^{4}.
\label{eq-fceelms2}
\end{equation}
% -----------------------------------------------
% In a standard way one finds non-zero tetrad components of the Einstein
% tensor
% -----------------------------------------------
% \begin{eqnarray*}
% G_{(0)(0)}=G_{(3)(3)} &=&
% \frac{4q^{2}\left[4(\varphi^{2}+z^{2})+r^{2}\right]}{r^{2}f(r,\varphi,z)^{4}},\\
% G_{(1)(1)}=-G_{(2)(2)} &=&
% \frac{4q^{2}\left[4(\varphi^{2}+z^{2})-r^{2}\right]}{r^{2}f(r,\varphi,z)^{4}},\\
% G_{(1)(2)}=G_{(2)(1)} &=&
% \frac{16q^{2}\sqrt{\varphi^{2}+z^{2}}}{rf(r,\varphi,z)^{4}}.
% \end{eqnarray*}
% -----------------------------------------------

The electromagnetic field is of magnetic type, since
% -----------------------------------------------
\begin{equation}
F_{(\mu )(\nu )}F^{(\mu )(\nu )} =
\frac{8q^{2}\left[ 4\left(\varphi^{2}/C^2+z^{2}\right) + r^{2}\right]}
{r^{2}f(r,\varphi,z)^{4}} \geq 0,
\label{eq-invelms2}
\end{equation}
% -----------------------------------------------
the tetrad components of the magnetic field strength read as
% -----------------------------------------------
\begin{equation}
F^{(1)(3)} = -B^{(2)} =
\frac{4q\sqrt{\varphi^{2}/C^2+z^{2}}}{rf(r,\varphi,z)^{2}},\quad
F^{(2)(3)} = B^{(1)} =
\frac{2q}{f(r,\varphi,z)^{2}}.
\label{eq-elmgelms2}
\end{equation}
% -----------------------------------------------
The physical interpretation of (\ref{eq-dselms2}) is rather ambiguous
because of the negative value of $\sigma$. As we have already mentioned in
section~\ref{sec-lc}, in case of negative $\sigma$ the problem of physical
cylindrical symmetric sources was not solved in a satisfactory way even for
the seed LC metric. Bonnor (see references in~\cite{bonnor}) has proved that
the seed LC metric for $\sigma=-1/2,\ C=1$ is locally isometric to Taub's
plane solution. Therefore, a more suitable alternative seems to be a point
of view preferred by Wang at all~\cite{wang}, that both the LC solution in
case of negative $\sigma$ and the metric (\ref{eq-dselms2}) are plane
symmetric.

The Kretschmann scalar
% -----------------------------------------------
\begin{equation}
\def\arraystretch{1.5}
\begin{array}{@{}*{4}{l}}
{\cal R} &=&\displaystyle
\frac{64}{f(r,\varphi,z)^{8}r^{12}}
\left. \rule{0pt}{3ex}\right[
3q^{8}r^{16}\left( \varphi^{2}/C^2+z^{2}\right)^{2}\times \\
&&\displaystyle\times\left(
21\varphi^{4}/C^4+42\varphi^{2}z^{2}/C^2+6\varphi^{2}r^{2}/C^2
             +6r^{2}z^{2}+21z^{4}+r^{4}\right)-
\\
&&\displaystyle -\,{}6q^{6}r^{12}\left(\varphi^{2}/C^2+z^{2}\right)\left( 
18\varphi^{4}/C^4+7\varphi^{2}r^{2}/C^2+36\varphi^{2}z^{2}/C^2+\right.\\
&&\displaystyle \left. +\,{}7r^{2}z^{2}+18z^{4}+r^{4}\right)+q^{4}r^{8}
\left(62z^{4}+124\varphi^{2}z^{2}/C^2+62\varphi^{4}/C^4+\right.\\
&&\displaystyle\left.+\,{}46\varphi^{2}r^{2}/C^2+46r^{2}z^{2}+5r^{4}\right)
+6q^{2}r^{4}\left( 2z^{2}-r^{2}+2\varphi^{2}/C^2\right)+3
\left.\rule{0pt}{3ex}\right].
\end{array}
\label{eq-Relms2}
\end{equation}
% -----------------------------------------------
diverges at $r=0$; this one-dimensional singularity along the
$z$-axis might indicate the location of the infinite line source.
The Petrov type is $I$ with Weyl scalars
% -----------------------------------------------
\refstepcounter{equation}
$$
\eqalignno{
\Psi_{0} = \Psi_{4} = & -12\,{}
\frac{q^{2}\left(\varphi^{2}/C^2+z^{2}\right)
\left[q^{2}r^{4}\left(\varphi^{2}/C^2+z^{2}\right)-1\right]}
{r^{2}f(r,\varphi,z)^{4}}
& (\theequation.{\rm a})\cr
\Psi_{1} = -\Psi_{3} = & -6\,{}
\frac{q^{2}\sqrt{\varphi^{2}/C^2+z^{2}}
\left[q^{2}r^{4}\left(\varphi^{2}/C^2+z^{2}\right)-1\right]}
{rf(r,\varphi,z)^{4}},
& (\theequation.{\rm b})\cr
\Psi_{2} = & \frac{2}{r^{6}f(r,\varphi,z)^{4}}
\left\{\left[q^{2}r^{4}\left(\varphi^{2}/C^2+z^{2}\right)-1\right]\right.
   \times &\cr
&\left.
\left[3q^{2}r^{4}\left(\varphi^{2}/C^2+z^{2}\right)-q^{2}r^{6}+1\right]
\right\}.
     & (\theequation.{\rm c})\cr}
$$\label{eq-weylscelms2}%
% -----------------------------------------------

% -----------------------------------------------
\section{The electrovacuum solution for $\sigma=1/4$}\nopagebreak
% -----------------------------------------------
\label{sec-elms4}

The last Killing vector listed in section~\ref{sec-lc} is $\bxi_{1/4}$ characterizing a
boost in $t-\varphi$ plane. As in the previous section, one should
preferably perform coordinate transformation
% -----------------------------------------------
$$
t=X\cosh Y,\qquad \varphi=CX\sinh Y
$$
% -----------------------------------------------
to simplify the calculus. This transformation does not map the whole
spacetime but only the interior of the light cone $t^2-\varphi^{2} > 0$. The
rest of the spacetime can be mapped analogously when we interchange
the hyperbolic functions and then one should formulate suitable boundary
conditions to couple those maps smoothly together. Here we shall restrict
ourselves only to the interior of the light cone. After substituting
$\sigma=1/4$ Eq.~(\ref{eq-dslc}) turns into
% -----------------------------------------------
$$
{\rm d}s^{2}=-\,r{\rm d}X^{2} + \frac{{\rm d}r^{2}}{\sqrt{r}}+
rX^{2}{\rm d}Y^{2}+\frac{{\rm d}z^{2}}{\sqrt{r}}.
$$
% -----------------------------------------------
It is worth mentioning that for $\sigma=1/4$ the LC metric
represents a transformation of one of the Kinnersley's type $D$
metric~\cite{kinnersley69} (his Case IVB with his $C=1$). For the
Killing vector one gets 
$\bxi_{1/4} = \partial_{Y} = rX^{2}\vec{dY}$. The choice of the tetrad
% -----------------------------------------------
\begin{equation}
\def\arraystretch{1.5}
\begin{array}{@{}*{4}{l}}
\displaystyle\bsmomega^{(0)}=\sqrt{r}F(X,r)\vec{dX},& \quad &
\displaystyle\bsmomega^{(1)}=\frac{F(X,r)}{r^{1/4}}\vec{dr},\cr
\displaystyle\bsmomega^{(2)}=\frac{\sqrt{r}X}{F(X,r)}\vec{dY},& \quad &
\displaystyle\bsmomega^{(3)}=\frac{F(X,r)}{r^{1/4}}\vec{dz}
\end{array}
\label{eq-baseelms4a}
\end{equation}
% -----------------------------------------------
gives the metric
% -----------------------------------------------
$$
{\rm d}s^{2}=-\,F(X,r)^2 r{\rm d}X^{2} + 
\frac{F(X,r)^{2}}{\sqrt{r}}{\rm d}r^{2}
+\frac{rX^{2}}{F(X,r)^{2}}{\rm d}Y^{2}+
\frac{F(X,r)^{2}}{\sqrt{r}}{\rm d}z^{2},
$$
% -----------------------------------------------
% and the solution of EM equations then
% -----------------------------------------------
% \begin{equation}
$$ F(X,r) = 1 + q^{2}X^{2}r;$$
% \label{eq-fceelms4a}
% \end{equation}
% -----------------------------------------------
and set the four-potential
% -----------------------------------------------
\begin{equation}
\vec{A} = q\frac{rX^{2}}{F(X,r)}\vec{dY}.
\label{eq-vpelms4a}
\end{equation}
% -----------------------------------------------
Returning back to the cylindrical coordinates one obtains the vector
potential
% -----------------------------------------------
\begin{equation}
\vec{A} = q\frac{r}{Cf(t,r,\varphi)}
\left(-\varphi{\vec dt}+t{\vec d}\bvphi\right)=
q\sqrt{r}\sqrt{t^{2}-\varphi^{2}/C^2}\bsmomega^{(2)},
\label{eq-vpelms4}
\end{equation}
% -----------------------------------------------
the basis tetrad
% -----------------------------------------------
\begin{equation}
\def\arraystretch{1.5}
\begin{array}{@{}*{4}{l}}
\displaystyle\bsmomega^{(0)}=\frac{\sqrt{r}f(t,r,\varphi)}{\sqrt{t^{2}
   -\varphi^{2}/C^2}}\left(t{\vec dt}-\frac{\varphi}{C^2}{\vec d}\bvphi\right),
   &\quad &
\displaystyle\bsmomega^{(1)}=\frac{f(t,r,\varphi)}{r^{1/4}}\vec{dr},\\
\displaystyle\bsmomega^{(2)}=\frac{\sqrt{r}}{\sqrt{t^{2}-\varphi^{2}/C^2}
   f(t,r,\varphi)}\left(-\frac{\varphi}{C}{\vec dt}+\frac{t}{C}{\vec d}\bvphi
                        \right),&\quad &
\displaystyle\bsmomega^{(3)}=\frac{f(t,r,\varphi)}{r^{1/4}}\vec{dz},
\end{array}   
\label{eq-baseelms4}
\end{equation}
% -----------------------------------------------
and the line element
% -----------------------------------------------
\begin{equation}
\def\arraystretch{1.5}
\begin{array}{@{}*{4}{l}}
{\rm d}s^{2}&=&\displaystyle\,{}
\frac{r}{t^{2}-\varphi^{2}/C^2}
\left[-\,f(t,r,\varphi)^{2}\left(t{\rm d}t-\frac{\varphi}{C^2}{\rm d}
\varphi\right)^{2}\right.+\\
&&\displaystyle\left.+\,{}
\frac{1}{f(t,r,\varphi)^{2}}\left(-\frac{\varphi}{C}{\rm d}t+
\frac{t}{C}{\rm d}\varphi\right)^{2}\right] +
\frac{f(t,r,\varphi)^{2}}{\sqrt{r}}
   \left({\rm d}r^{2}+{\rm d}z^{2}\right),
\end{array}
\label{eq-dselms4}
\end{equation}
% -----------------------------------------------
where
% -----------------------------------------------
\begin{equation}
f(t,r,\varphi) = 1 + q^{2}\left(t^{2}-\varphi^{2}/C^2\right)r.
\label{eq-fceelms4}
\end{equation}
% -----------------------------------------------
% and tetrad components of the Einstein tensor
% -----------------------------------------------
% \begin{eqnarray*}
% G_{(0)(0)}=G_{(1)(1)} &=&
% \frac{q^{2}\left(4\sqrt{r}+t^{2}-\varphi^{2}\right)}
%      {\sqrt{r}f(t,r,\varphi)^{4}},\\
% G_{(2)(2)}=-G_{(3)(3)} &=&
% \frac{q^{2}\left(-4\sqrt{r}+t^{2}-\varphi^{2}\right)}
%      {\sqrt{r}f(t,r,\varphi)^{4}},\\
% G_{(0)(1)}=G_{(1)(0)} &=&
% -\,\frac{4q^{2}\sqrt{t^{2}-\varphi^{2}}}{r^{1/4}f(t,r,\varphi)^{4}}.
% \end{eqnarray*}
% -----------------------------------------------

This time the electromagnetic field is neither purely electric, nor
purely magnetic but its type is different at various places and at various
time, which can be demonstrated by the invariant
% -----------------------------------------------
\begin{equation}
F_{(\mu )(\nu )}F^{(\mu )(\nu )} =
-\,\frac{2q^{2}\left[ 4\sqrt{r}-\left(t^{2}-\varphi^{2}/C^2\right)\right]}
{\sqrt{r}f(t,r,\varphi)^{4}}
\label{eq-invelms4}
\end{equation}
% -----------------------------------------------
and components of the electromagnetic field tensor
% -----------------------------------------------
\begin{equation}
F^{(0)(2)} = E^{(2)} =
\frac{2q}{f(t,r,\varphi)^{2}},\quad
F^{(1)(2)} = B^{(3)} =
\frac{q\sqrt{t^{2}-\varphi^{2}}}{r^{1/4}f(t,r,\varphi)^{2}}.
\label{eq-elmgelms4}
\end{equation}
% -----------------------------------------------
This rather strange behaviour originates in the fact that the tetrad
(\ref{eq-baseelms4}) is carried by an observer moving round the
infinite line source in azimuthal direction, that means, rotating round
$z$-axis. In this sense the indefinite character of electromagnetic field
can be understood as a special-relativistic effect. The EM field
(\ref{eq-dselms4}) can be interpreted as an infinite line source with linear
density $2.5\cdot 10^{27}\,\mbox{g}\cdot\mbox{cm}^{-1}$ in the external
electromagnetic field, part of which is analogous to the Bonnor-Melvin
longitudinal magnetic background. The solution (\ref{eq-dselms4}) is
non-stationary and in contrast to the seed LC metric it is neither
cylindrically, nor axially symmetric.

In the interior of the light cone $t^{2}-\varphi^{2}/C^2 > 0$ the metric
(\ref{eq-dselms4}) has again one-dimensional singularity at $r=0$
where the Kretschmann scalar
% -----------------------------------------------
\begin{equation}
\def\arraystretch{1.5}
\begin{array}{@{}*{4}{l}}
{\cal R}\!&\!=\!&\!\displaystyle
\frac{1}{4f(t,r,\varphi)^{8}r^{7/2}}
\left. \rule{0pt}{3ex}\right[
3q^{8}\left(t^{2}-\varphi^{2}/C^2\right)
\left(21\varphi^{4}r^{9/2}/C^4-\right.\cr
&&\displaystyle\left.-\,{}42\varphi^{2}t^{2}r^{9/2}/C^2+
96\varphi^{2}r^{5}/C^2+256r^{11/2}+21t^{4}r^{9/2}-96t^{2}r^{5}\right)-\cr
&&\displaystyle-\,{}12q^{6}\left(t^{2}-\varphi^{2}/C^2\right)
\left( 9\varphi^{4}r^{7/2}/C^4+56\varphi^{2}r^{4}/C^2
      -18\varphi^{2}t^{2}r^{7/2}/C^2+\right.\cr
&&\displaystyle\left.+\,{}9t^{4}r^{7/2}-56r^{4}t^{2}+128r^{9/2}\right)
+2q^{4}\left(368\varphi^{2}r^{3}/C^2-368r^{3}t^{2}+\right.\cr
&&\displaystyle\left.+\,{}31\varphi^{4}r^{5/2}/C^4+31r^{5/2}t^{4}+
640r^{7/2}-62\varphi^{2}r^{5/2}t^{2}/C^2\right)+\\
&&\displaystyle +\,{}12q^{2}
\left( t^{2}r^{3/2}-\varphi^{2}r^{3/2}/C^2+8r^{2}\right)+3\sqrt{r}
\left.\rule{0pt}{3ex}\right].
\end{array}
\label{eq-Relms4}
\end{equation}
% -----------------------------------------------
becomes infinite. The Petrov type is $I$ with the Weyl scalars
% -----------------------------------------------
\refstepcounter{equation}
$$
\eqalignno{
\Psi_{0} = \Psi_{4} = & \frac{3}{4}\,
\frac{q^{2}\left( t^{2}-\varphi^{2}/C^2\right)
\left[q^{2}r\left(t^{2}-\varphi^{2}/C^2\right)-1\right]}
{\sqrt{r}f(t,r,\varphi)^{4}},
& (\theequation.{\rm a})\cr
\Psi_{1} = -\Psi_{3} = & -\,\frac{3q^{2}}{2}
\frac{q^{2}r\left(t^{2}-\varphi^{2}/C^2\right)^{2}
   -\left(t^{2}-\varphi^{2}/C^2\right)}
{r^{1/4}\sqrt{t^{2}-\varphi^{2}/C^2}f(t,r,\varphi)^{4}},
& (\theequation.{\rm b})\cr
\Psi_{2} = & \frac{1}{8r^{3/2}f(t,r,\varphi)^{4}}
\left[3q^{4}r^{2}\left(t^{2}-\varphi^{2}/C^2\right)^{2}+
16q^{4}r^{5/2}\left(t^{2}-\varphi^{2}/C^2\right)-\right.\cr
& \left.-\,{}16q^{2}r^{3/2}-2q^{2}r\left(t^{2}-\varphi^{2}/C^2\right)-1
\right].
     & (\theequation.{\rm c})\cr}
$$\label{eq-weylscelms4}%
% -----------------------------------------------

% -----------------------------------------------
\section{The LC solution with longitudinal electric field}\nopagebreak
% -----------------------------------------------
\label{sec-elwz}

In preceding sections all Killing vectors of the LC metric were exhausted to
generate EM field. As a by-product of those calculations one more EM field
was found, or surprisingly, new possible interpretation was assigned to the
metric (\ref{eq-dsmgz}) derived in section~\ref{sec-mgwz}. It should be
pointed out that though the next steps and considerations follow the scheme
of the HM conjecture described above, there is a key difference: we do not
employ any Killing vector of the seed LC metric as a vector four-potential.
On the other hand, one should emphasize that this fact does not contradict
the conjecture, which has never been considered as the only possibility of
generating EM fields.

Let us take the boost vector potential
% -----------------------------------------------
\begin{equation}
\vec{A} = q\left(z\vec{dt}-t\vec{dz}\right) =
\frac{q}{f(r)}\left(zr^{-2\sigma}\bsmomega^{(0)}-tr^{2\sigma}\bsmomega^{(3)}
\right)
\label{eq-vpelwz}
\end{equation}
% -----------------------------------------------
with tetrad (\ref{eq-basemgz}) inducing metric (\ref{eq-dsmgz}) and
with the function $f(r)$ in the form (\ref{eq-fcemgz}).
% -----------------------------------------------
% \begin{equation}
% f(r) = 1 + c_{1}r^{2(1 - 2\sigma )}.
% \label{eq-fceelwz}
% \end{equation}
% -----------------------------------------------
Solving EM equations one derives the constant $c_{1}$
% -----------------------------------------------
\[
c_{1}=\frac{q^{2}}{(2\sigma -1)^{2}}.
\]
% -----------------------------------------------
% and components of the Einstein tensor
% -----------------------------------------------
% \begin{equation}
% G_{(0)(0)}=G_{(1)(1)}=G_{(2)(2)}=-\,G_{(3)(3)} =
% \frac{4q^{2}}{r^{8\sigma^{2}}f(r)^{4}}.
% \label{eq-einstelwz}
% \end{equation}
% -----------------------------------------------

The electromagnetic invariant
% -----------------------------------------------
\begin{equation}
F_{(\mu )(\nu )}F^{(\mu )(\nu )} =
-\,\frac{8q^{2}C^{2}\left( 2\sigma-1\right)^{2}}
{r^{8\sigma^{2}}f(r)^{4}} < 0,
\label{eq-invelwz}
\end{equation}
% -----------------------------------------------
so the electromagnetic field represents an electric type with
longitudinal electric field strength oriented along the $z$-axis
% -----------------------------------------------
\begin{equation}
F^{(0)(3)} = E^{(3)} =
-\,\frac{2q}{f(r)^{2}r^{4\sigma^{2}}}.
\label{eq-elmgelwz}
\end{equation}
% -----------------------------------------------
Thus we have generated EM field of an infinite line source in a universe
filled with the longitudinal electric field. The source of the electric
field can be hardly identified with the charge distribution along the
$z$-axis. Therefore the electric field should be understood like a
background which has the same role as the magnetic field in the
Bonnor-Melvin universe. Moreover, substituting $\sigma=0,\ C=1$ one is left
with the metric (\ref{eq-dsbonmel}), the Bonnor-Melvin universe. The
conclusion is straightforward: the Bonnor-Melvin universe need not necessary
represent a magnetic EM field but also an electric one or their
combination.

The Kretschmann scalar
% -----------------------------------------------
\begin{equation}
\begin{array}{@{}*{4}{l}}
{\cal R} &=&\displaystyle
\frac{64\left(2\sigma -1\right)^{2}}{f(r)^{8}r^{8\sigma (2\sigma -1)+4}}
\left. \rule{0pt}{3ex}\right[g_{4}(r)^{4}
\left(\sigma-1\right)^{2}\left( 4\sigma^{2}-6\sigma + 3 \right)+\\
&&\displaystyle +\,{}6 g_{4}(r)^{3}
\left( 2\sigma-1\right)\left( \sigma-1\right)^{2}-\\
&&\displaystyle-\,{}g_{4}(r)^{2}
\left(8\sigma^{4}-16\sigma^{3}-12\sigma^{2}+20\sigma-5\right) -\\
&&\displaystyle -\,{}6g_{4}(r)
\sigma\left(2\sigma-1\right)+
\,{}\sigma^{2}\left(4\sigma^{2}-2\sigma+1
\right)\left.\rule{0pt}{3ex}\right],
\end{array}
\label{eq-Relwz}
\end{equation}
% -----------------------------------------------
where $g_{4}(r)= q^{2}r^{2-4\sigma }/\left(2\sigma-1\right)^{2}$ is again
singular at $r=0$ with the exception $\sigma=1/2$, in which case the spacetime
is flat and does not include any electric field.

Analogously to (\ref{eq-dsmgz}) the metric is Petrov type $I$ with
Weyl scalars
% -----------------------------------------------
\refstepcounter{equation}
$$
\eqalignno{
\Psi_{0} = & \Psi_{4} = - \frac{q^{2}r^{2}-r^{4\sigma}\left(2\sigma
-1\right)^{2}}{r^{8\sigma^{2}+16\sigma-10}(2\sigma -1)^{3}f(r)^{4}}
\left[ r^{4\sigma }\left(8\sigma^{4}-4\sigma^{3}-2\sigma^{2}+
                         \sigma\right)+\right. &\cr
&\hspace{5ex}\left.+\,{}q^{2}r^{2}\left(2\sigma^{2}-5\sigma+3\right)\right],
& (\theequation.{\rm a})\cr
\Psi_{2} = & - \frac{q^{2}r^{2}-r^{4\sigma}\left(2\sigma
-1\right)^{2}}{r^{8\sigma^{2}+16\sigma-10}(2\sigma -1)^{2}f(r)^{4}}
\left[
r^{4\sigma}\sigma\left(2\sigma -1\right)^{2}
+ q^{2}r^{2}\left(\sigma -1\right) \right],
     & (\theequation.{\rm b})\cr}
$$\label{eq-weylscelwz}%
% -----------------------------------------------
algebraic special cases are summarized in the following table:
% -----------------------------------------------
% \begin{table}[h]
% \caption{Petrov types for special values of the parameter
%    $\sigma$ (The LC solution with longitudinal electric field).}
% \label{tab-elwz}
\begin{center}
\renewcommand{\arraystretch}{2}
\begin{tabular}{|c|c|c|}
\hline
$\sigma$ & Petrov type & Weyl Scalars\\
\hline
0 & $D$ &
$\Psi_{0} = \Psi_{4} = 3\Psi_{2} =
\displaystyle{\frac{3q^{2}\left( qr+1\right)\left( qr -1\right)}
{\left(q^{2}r^{2}+1\right)^{4}}}$\\
$\displaystyle{1\over 2}$ & 0 & all zero\\
1 & $D$ &
$\Psi_{0} = \Psi_{4} = 3\Psi_{2} =
\displaystyle{\frac{3\left( r^{2}-q^{2}\right)}
{\left(r^{2}+q^{2}\right)^{4}}}$
\\[1.5ex] \hline
\end{tabular}
\end{center}
% \end{table}
% -----------------------------------------------

% -----------------------------------------------
\section{Conclusion}\nopagebreak
% -----------------------------------------------

The application of the HM conjecture to the LC seed metric revealed several
interesting features:
% -----------------------------------------------
\begin{enumerate}
\item[(i)]
All the Killing vectors of the seed vacuum solution were employed to obtain
electromagnetic fields of both electric and magnetic type using HM
conjecture. The process of generation is marked by common algorithmic steps
(though not applicable generally) allowing to devise an instructive
scheme (section~\ref{sec-hmc}). All obtained solutions describe the EM
fields of an infinite line source (``endless wire'') with electromagnetic
field, the source of which can be either identified with a charge
(\ref{eq-dselw}), or with current (\ref{eq-dsmgph}) of the linear source
or with some electromagnetic background (\ref{eq-dsmgz}).
\item[(ii)]
The cylindrical symmetric EM fields 20.9a and 20.9b in~\cite{KSMH} do not
represent the most general cases since they do not include solutions
(\ref{eq-dsmgph}), (\ref{eq-dsmgz}).
\item[(iii)]
The Bonnor-Melvin universe (\ref{eq-dsbonmel}) need not necessarily contain
a longitudinal magnetic EM field as is usually supposed. According to the
results of section~\ref{sec-elwz} the background field might be both
electric and magnetic. Thus, using the HM conjecture we have obtained
a qualitatively new interpretation of the Bonnor-Melvin universe.
\item[(iv)]
Setting $\sigma=-1/2$ and following the transformation found by
Bonnor~\cite{bonnor} one can reduce each of the generated spacetimes (with
the exception~(\ref{eq-dselms4})) to some plane symmetric solution of the
Einstein-Maxwell equations. Thus we have obtained also Einstein-Maxwell
fields of Taub's type (either electric or magnetic ones) as special cases of
found solutions.
\end{enumerate}

The analysis of radial geodesic motion in the~appendix supports the above
interpretation of one-dimensional singularity located along the $z$-axis.
The singularity has an \emph{attractive} character that can be explained
naturally by the presence of an infinite line source, the character of which
is described in~\cite{wang}. In comparison with the LC seed metric the
presence of electromagnetic field generally results in a stronger
singularity's attraction.
% -----------------------------------------------

% -----------------------------------------------
\section*{Acknowledgement}\nopagebreak
% -----------------------------------------------

Authors are indebted to Prof. V. Majern\'{\i}k for stimulating comments and
fruitful discussions.

% ------------------------------------------------------------
\appendix
% ------------------------------------------------------------
\section{Appendix: radial geodesic motion}\nopagebreak

This part concentrates only on the static cylindrically symmetric
cases, namely the metrics~(\ref{eq-dslc}), (\ref{eq-dsmgph}),
(\ref{eq-dsmgz}) and (\ref{eq-dselw}) treated in sections~\ref{sec-lc},
\ref{sec-mgwaz}, \ref{sec-mgwz} and \ref{sec-elw} respectively and on the
solution derived in the section~\ref{sec-elwz}. In all these cases the
metric coefficients depend only on the radial distance from the $z$-axis
$r$, so one has three integrals of motion at his disposal: the covariant
coordinate components of particles four-velocity $u_{t}, u_{\varphi}, u_{z}$
defined in the common way (here we use the fact that all the metrics are
diagonal)
% -----------------------------------------------
\begin{displaymath}
u_{t}=g_{tt}\frac{{\rm d}t}{{\rm d}\tau},\quad
u_{\varphi}=g_{\varphi\varphi}\frac{{\rm d}\varphi}{{\rm d}\tau},\quad
u_{z}=g_{zz}\frac{{\rm d}z}{{\rm d}\tau},
\end{displaymath}
% -----------------------------------------------
where $\tau$ is the particle's proper time. The normalization
condition
% -----------------------------------------------
\begin{equation}
\label{eq-fourv}
u_{\mu}u^{\mu} = g^{tt}u_{t}^{2} + g_{rr}(u^{r})^{2} +
g^{\varphi\varphi}u_{\varphi}^{2} + g^{zz}u_{z}^{2} = -1
\end{equation}
% -----------------------------------------------
enables to express the square of the contravariant radial
four-velocity component
% -----------------------------------------------
\begin{equation}
\label{eq-ur2}
(u^{r})^{2}=\left(\frac{{\rm d}r}{{\rm d}\tau}\right)^{2} =
\frac{1}{g_{rr}}\left(-1 - g^{tt}u_{t}^{2} - 
g^{\varphi\varphi}u_{\varphi}^{2} - g^{zz}u_{z}^{2}\right).
\end{equation}
% -----------------------------------------------

The purely radial motion (there is, evidently, no dragging effect) is set by
putting $u_{\varphi}=u_{z}=0$.  Here, unfortunately, it is not possible to
introduce an effective potential independent of the particle energy per unit
mass $-u_{0}$.  Therefore, we use the absolute value of radial velocity
instead. The scale of the radial velocity is not important for qualitative
discussion, and thus it is not explicitly introduced in the drawings. The
boundary between the region with zero and non-zero radial velocity
physically determines the turning points for radial motion. To detect the
position of the turning points more exactly, the contour lines are drawn in
the base plane of each figure. The top flat part of the plots corresponds to
regions with high radial velocities.

Each figure is related to a different spacetime and includes four subplots.
These subplots correspond with two different values of particle energy
(first and second row of subplots), and with weaker or stronger electromagnetic
field (left and right column respectively). In this way we can illustrate
the influence of the electromagnetic field on radial geodesics and compare
the situation with motion in the seed LC metric. The first value of energy
($u_{0}=1$) characterizes a particle at rest in Minkowski spacetime.

\begin{figure}[ht!]
\parbox[b]{7cm}{
\centerline{
\hbox{
   \epsfxsize=7cm
   \rotatebox{-90}{\epsfbox{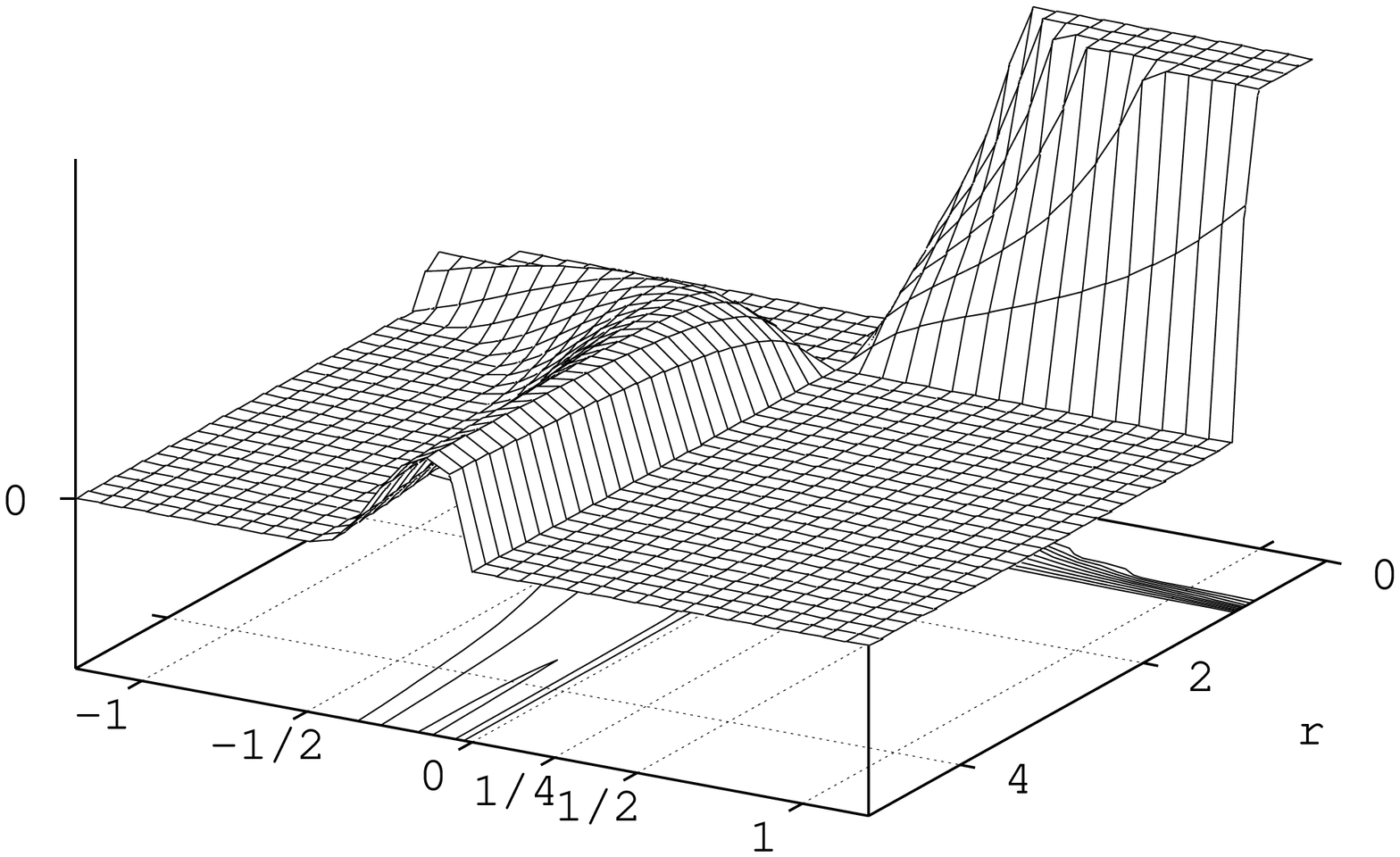}}}}

\centerline{(a)}
}
\hfill
\parbox[b]{7cm}{
\centerline{
\hbox{
   \epsfxsize=7cm
   \rotatebox{-90}{\epsfbox{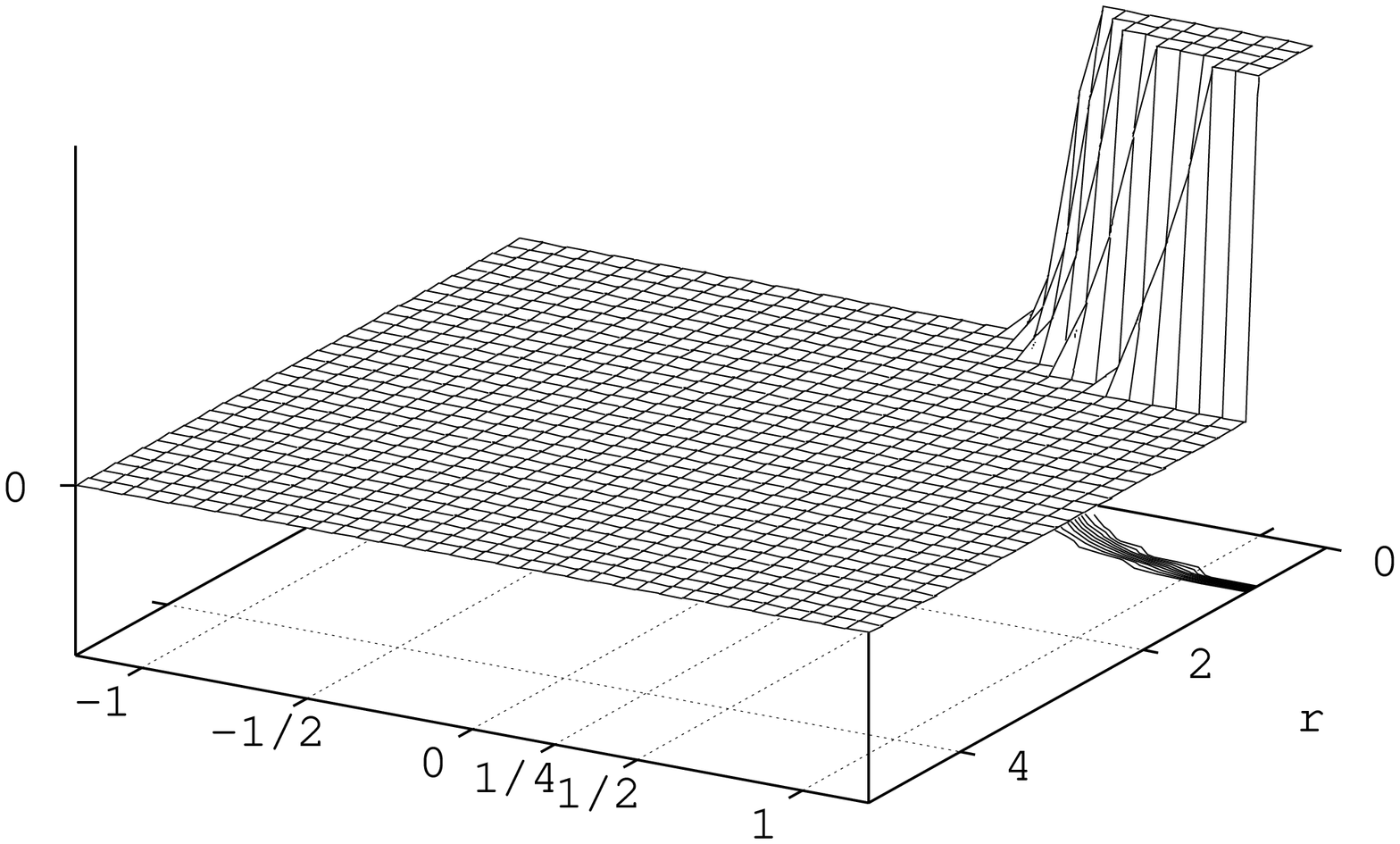}}}}

\centerline{(b)}
}

\vspace{5ex}
% -----------------------------------------------
\parbox[b]{7cm}{
\centerline{
\hbox{
   \epsfxsize=7cm
   \rotatebox{-90}{\epsfbox{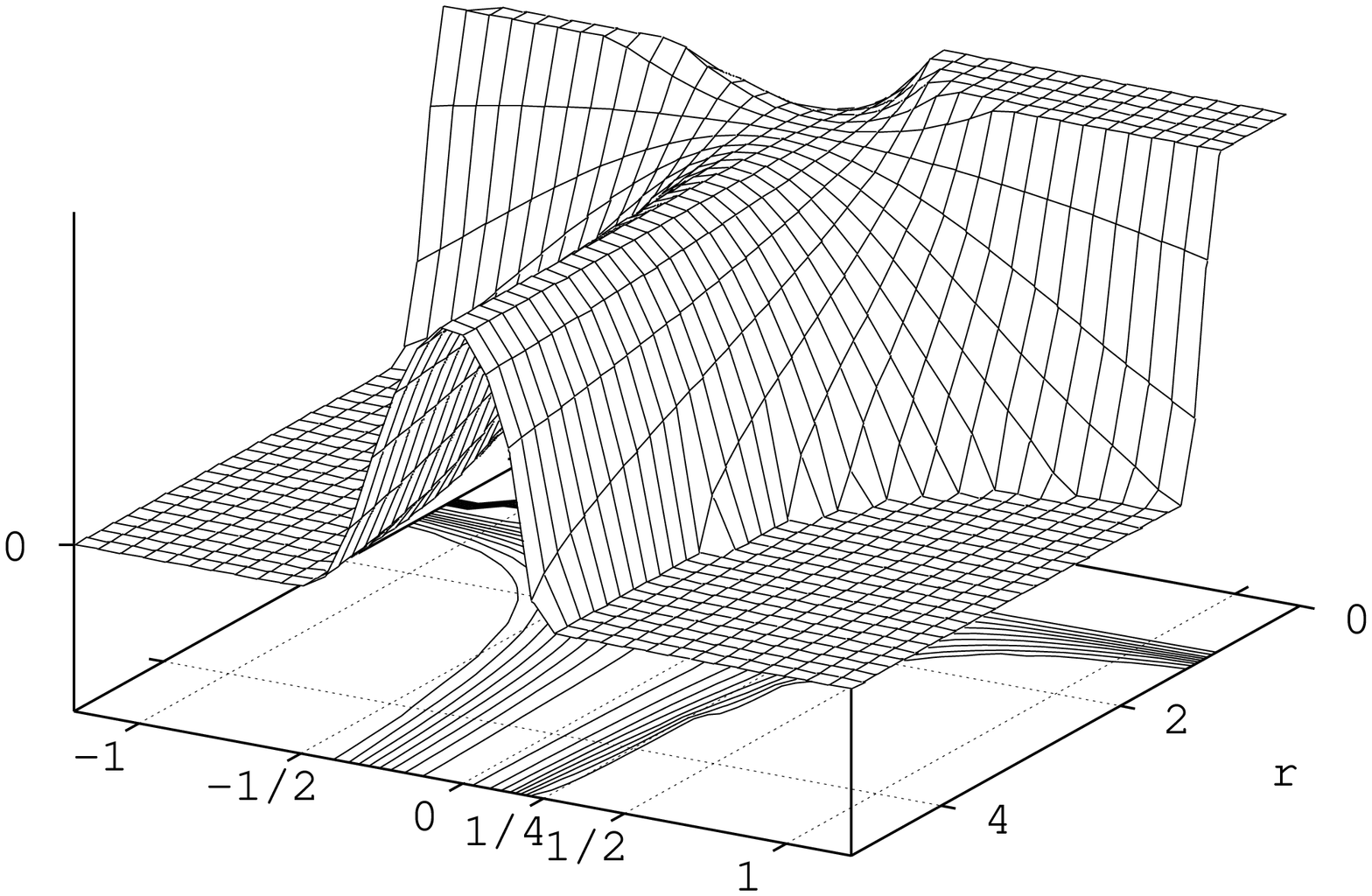}}}}

\centerline{(c)}
}
\hfill
\parbox[b]{7cm}{
\centerline{
\hbox{
   \epsfxsize=7cm
   \rotatebox{-90}{\epsfbox{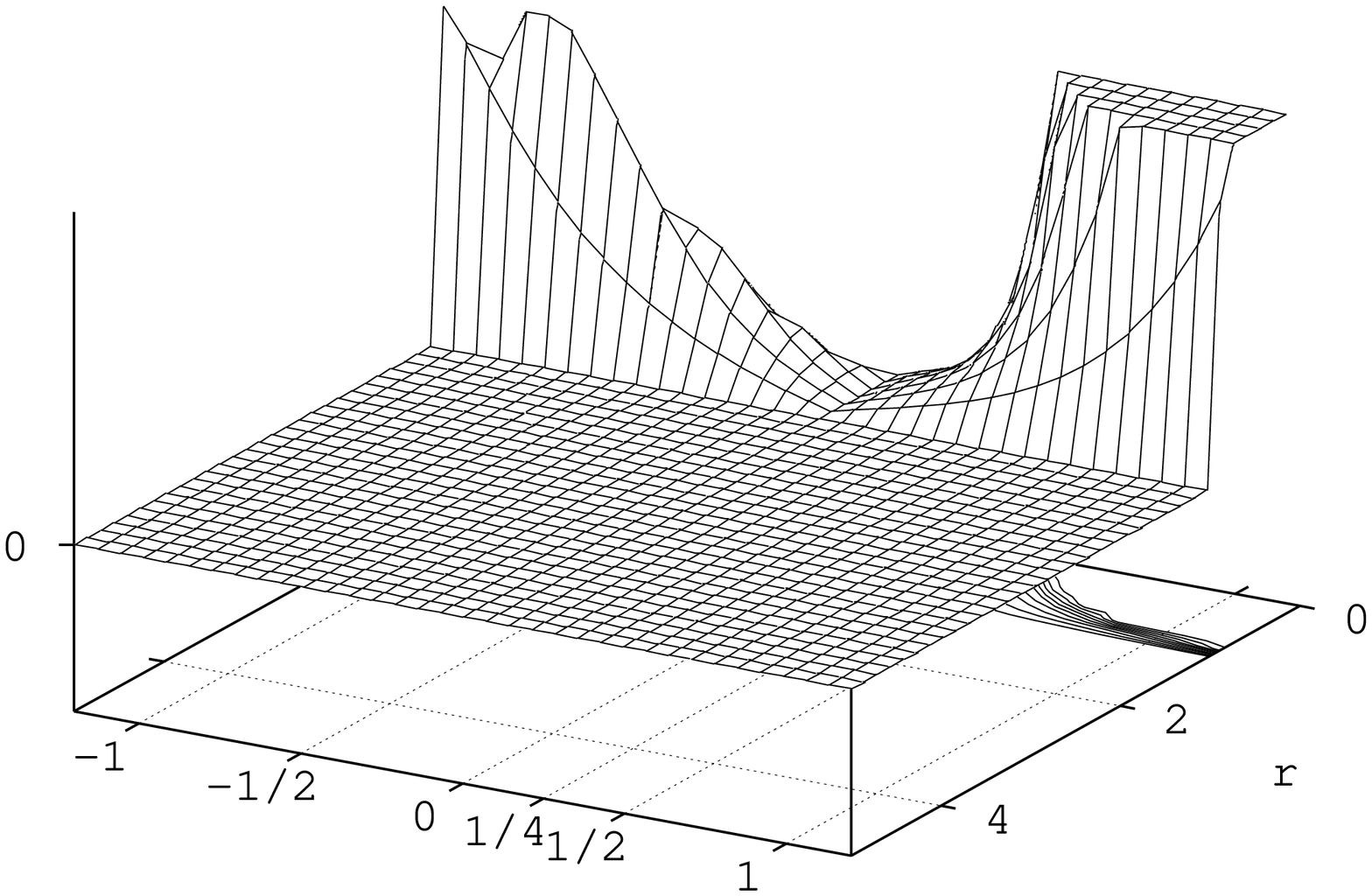}}}}

\centerline{(d)}
}
\caption{Absolute value of radial velocity for 
     the solution with azimutal magnetic field.
 (a) $u_{0}=1,\ q=0.1$.
 (b) $u_{0}=1,\ q=1$.
 (c) $u_{0}=2,\ q=0.1$.
 (d) $u_{0}=2,\ q=1$.
 }
\label{fig-mgph}
\end{figure}

The plots in Fig.~\ref{fig-mgph} represent the dependence of the
$|u^{r}|$ on $r$ for various values of $\sigma$, that means, for different
spacetimes from the class of solutions (\ref{eq-dsmgph}).  When the
spacetime is flat ($\sigma=0$) then the radial motion with constant energy
must result in constant radial velocity which equals zero in cases (a),
(b) and is non-zero in cases (c),(d) (the top of the ``ridge'').  While for
negative $\sigma$ (which is probably not relevant to any real situation)
the singularity is not attractive and particle is kept at some distance
from the $z$-axis, for $\sigma >0$ the radial velocity rapidly increases
towards the singularity with an evident attractive effect. The
Figs.~\ref{fig-mgph}(a) and (c) illustrating the situation in presence of
weak magnetic field are qualitatively identical to appropriate plots for
the seed LC solution~(\ref{eq-dslc}).

\begin{figure}[ht!]
\parbox[b]{7cm}{
\centerline{
\hbox{
   \epsfxsize=7cm
   \rotatebox{-90}{\epsfbox{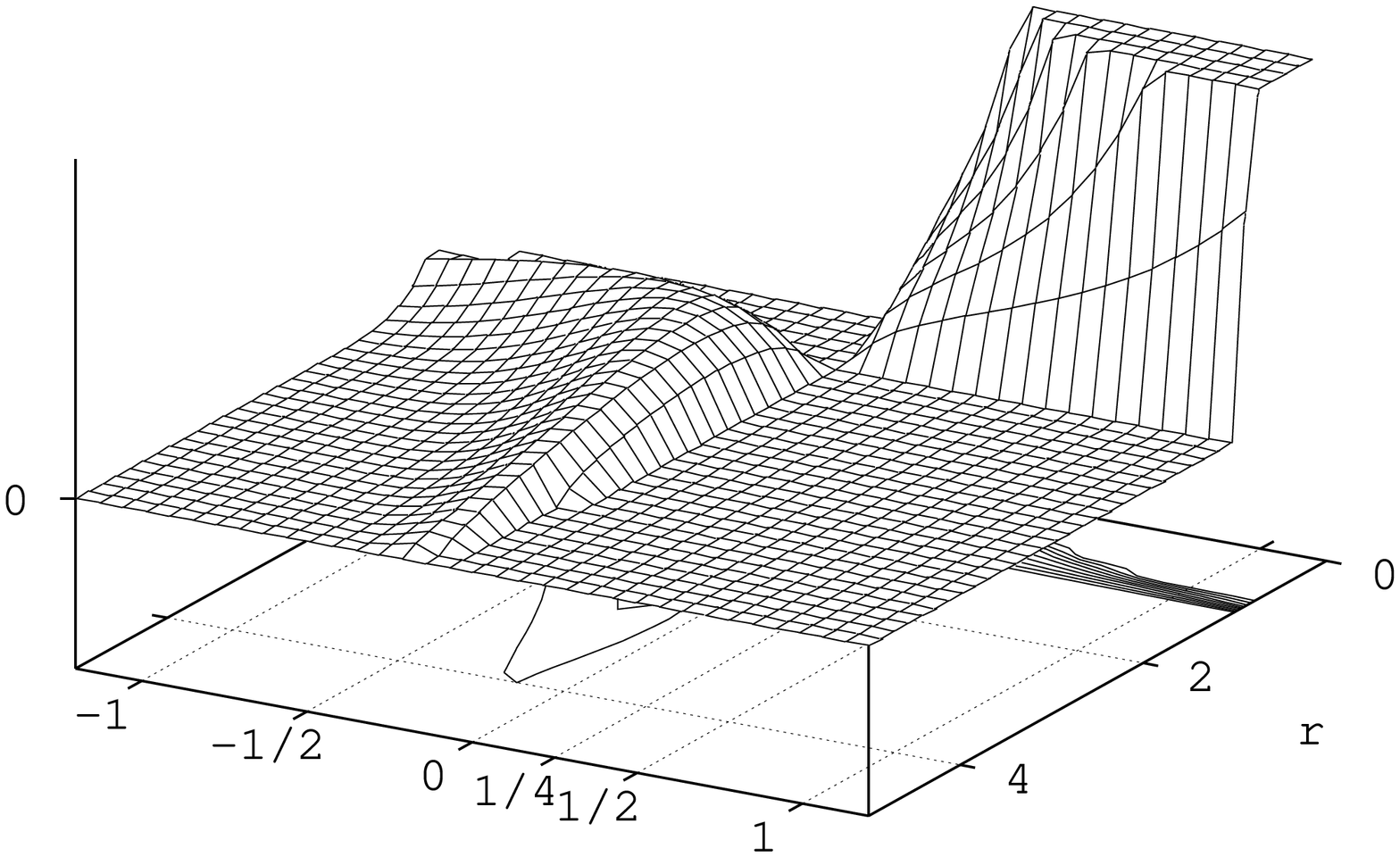}}}}

\centerline{(a)}
}
\hfill
\parbox[b]{7cm}{
\centerline{
\hbox{
   \epsfxsize=7cm
   \rotatebox{-90}{\epsfbox{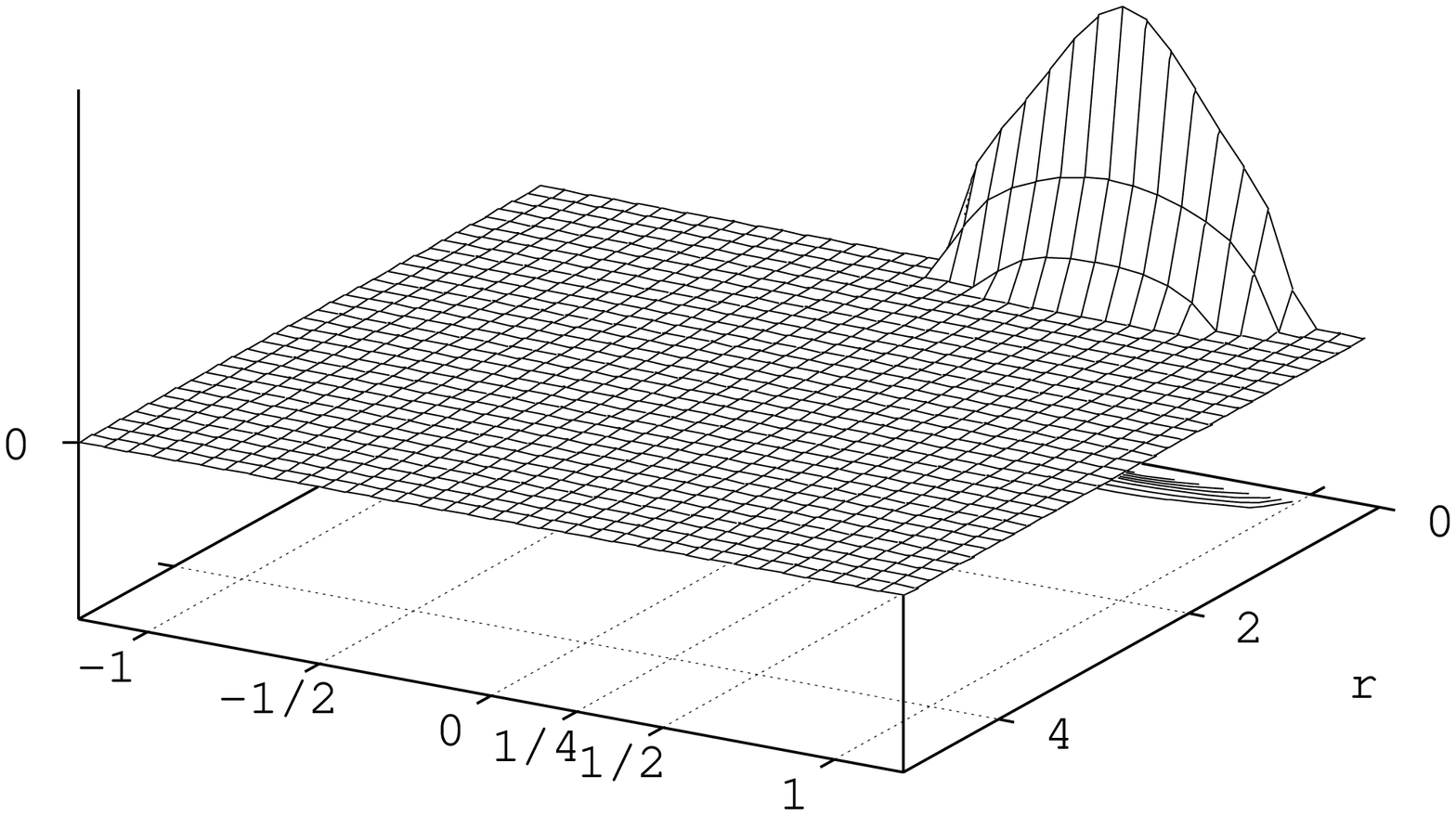}}}}

\centerline{(b)}
}

\vspace{5ex}
% -----------------------------------------------
\parbox[b]{7cm}{
\centerline{
\hbox{
   \epsfxsize=7cm
   \rotatebox{-90}{\epsfbox{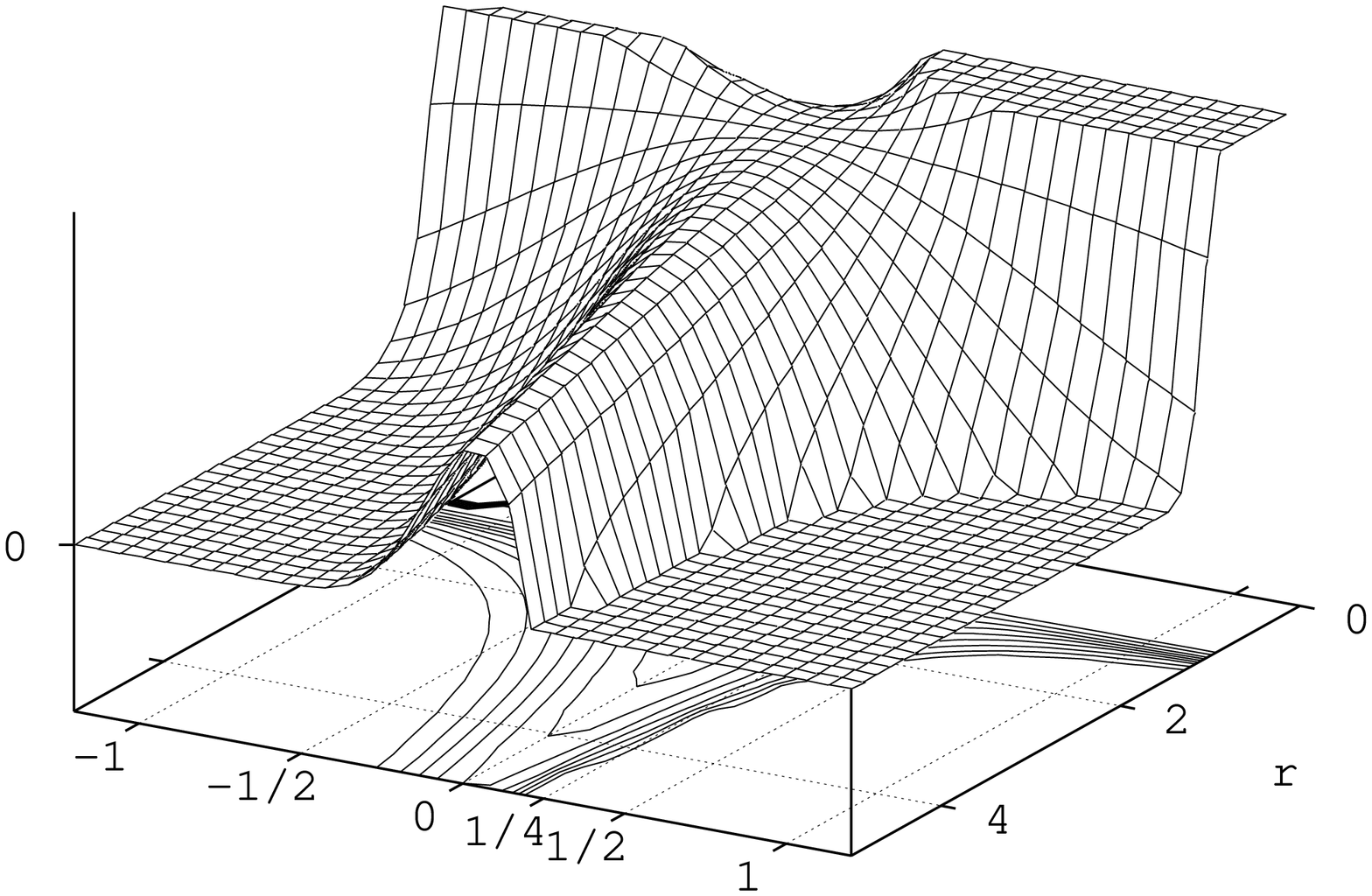}}}}

\centerline{(c)}
}
\hfill
\parbox[b]{7cm}{
\centerline{
\hbox{
   \epsfxsize=7cm
   \rotatebox{-90}{\epsfbox{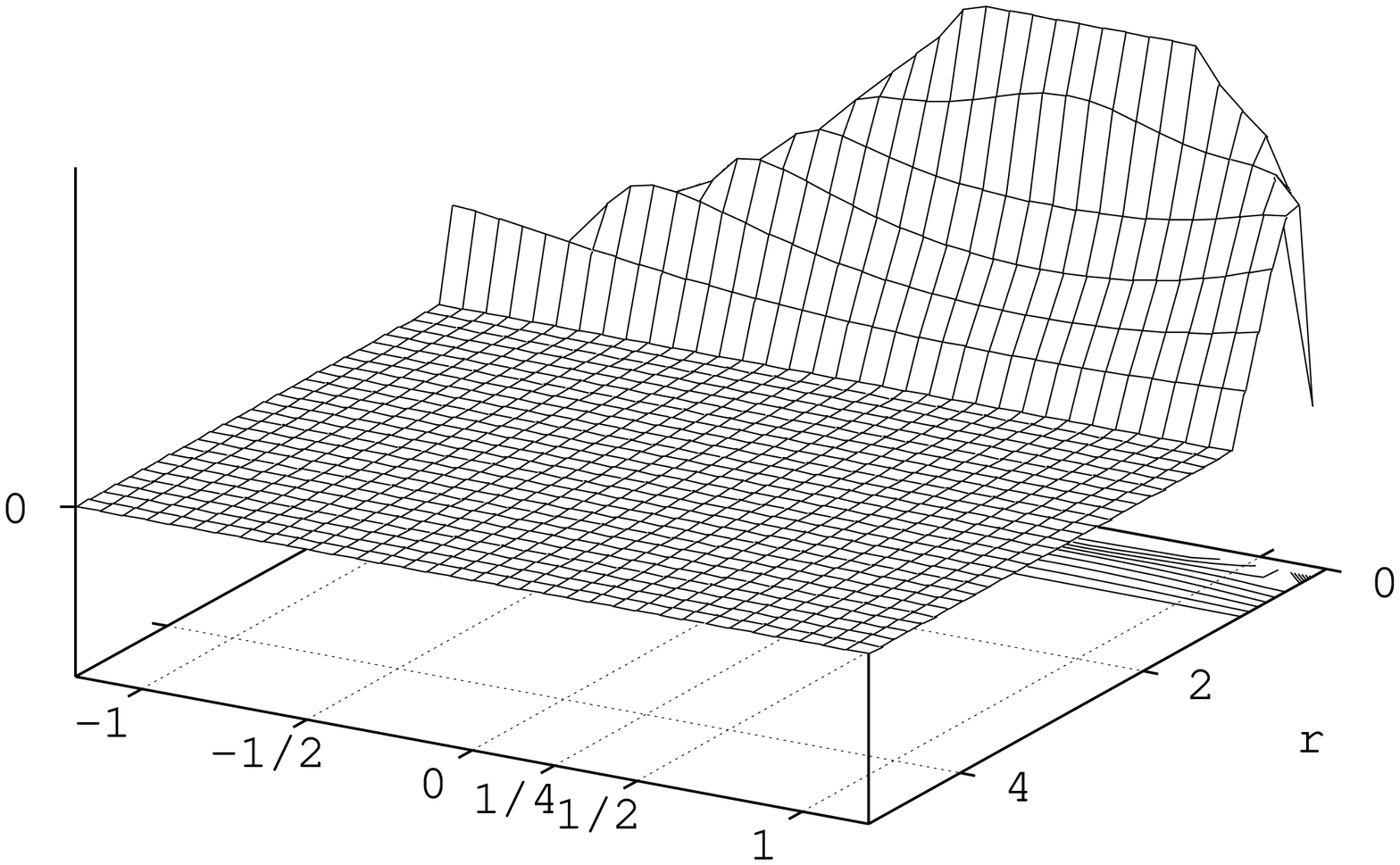}}}}

\centerline{(d)}
}
\caption{Absolute value of radial velocity for 
     the solution with longitudinal magnetic field.
 (a) $u_{0}=1,\ q=0.1$.
 (b) $u_{0}=1,\ q=1$.
 (c) $u_{0}=2,\ q=0.1$.
 (d) $u_{0}=2,\ q=1$.	
 }
\label{fig-mgz}
\end{figure}

Quite an analogous situation can be found in Fig.~\ref{fig-mgz}
belonging to the solution~(\ref{eq-dsmgz}) with longitudinal magnetic field
which degenerates to flat spacetime for $\sigma = 1/2$.  Apparently, one
should expect that in this case we again recognize the motion with a
constant radial velocity in the plots. There is, however, an important
difference originating in the form of the seed LC metric. Evidently, the
metric~(\ref{eq-dslc}) for $\sigma=1/2$ turns into
% -----------------------------------------------
\begin{displaymath}
{\rm d}s^{2}=-r^{2}{\rm d}t^{2} + {\rm d}r^{2} + {\rm d}\varphi^{2}
+ {\rm d}z^{2}
\end{displaymath}
% -----------------------------------------------
with interchanged components $g_{tt}$ and $g_{\varphi\varphi}$ compared to
the Minkowski spacetime; it rather corresponds to the frame of an
accelerated observer. Such an observer certainly will not measure a
constant radial velocity for a considered radial motion. On the
other hand, comparing Figs.~\ref{fig-mgph} and \ref{fig-mgz}, subplots (a),
(c), we can see there is no crucial difference between the motion in a
weak azimuthal and longitudinal magnetic field, the positions of turning
points nearly coincide. Some differences can be detected in stronger fields
(Figs.~\ref{fig-mgph} and \ref{fig-mgz}, subplots (b), (d)).

\begin{figure}[ht!]
\parbox[b]{7cm}{
\centerline{
\hbox{
   \epsfxsize=7cm
   \rotatebox{-90}{\epsfbox{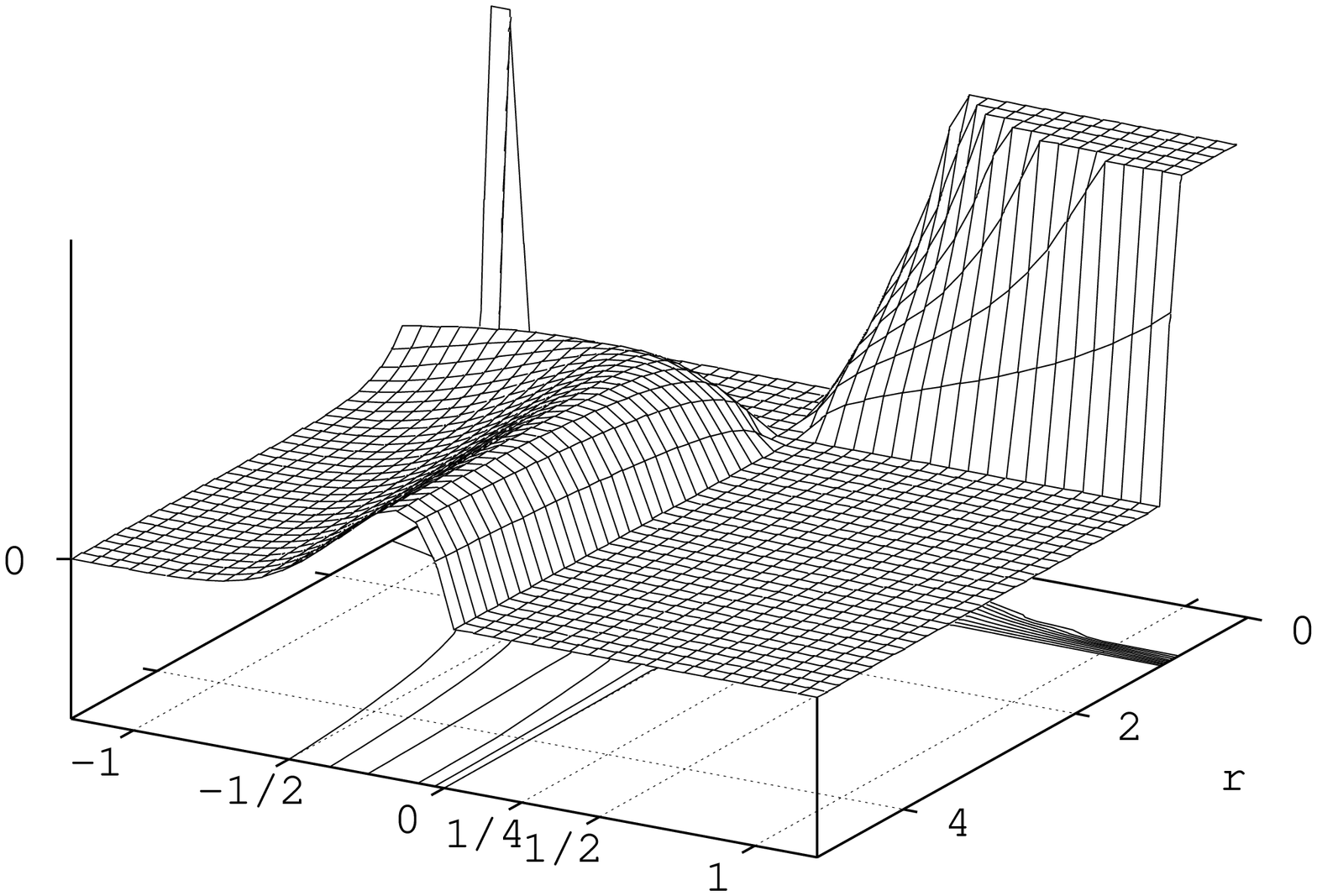}}}}

\centerline{(a)}
}
\hfill
\parbox[b]{7cm}{
\centerline{
\hbox{
   \epsfxsize=7cm
   \rotatebox{-90}{\epsfbox{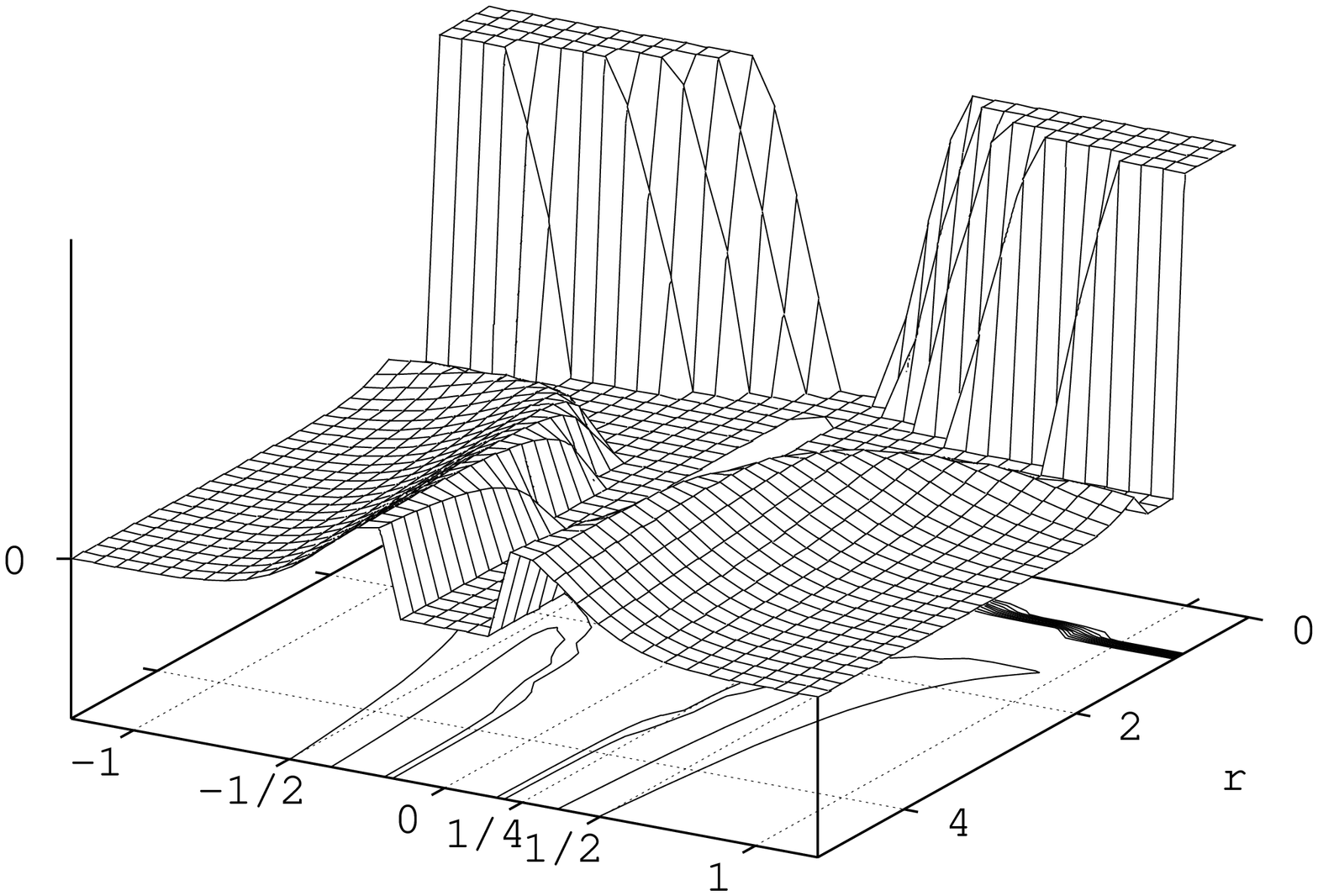}}}}

\centerline{(b)}
}

\vspace{5ex}
% -----------------------------------------------
\parbox[b]{7cm}{
\centerline{
\hbox{
   \epsfxsize=7cm
   \rotatebox{-90}{\epsfbox{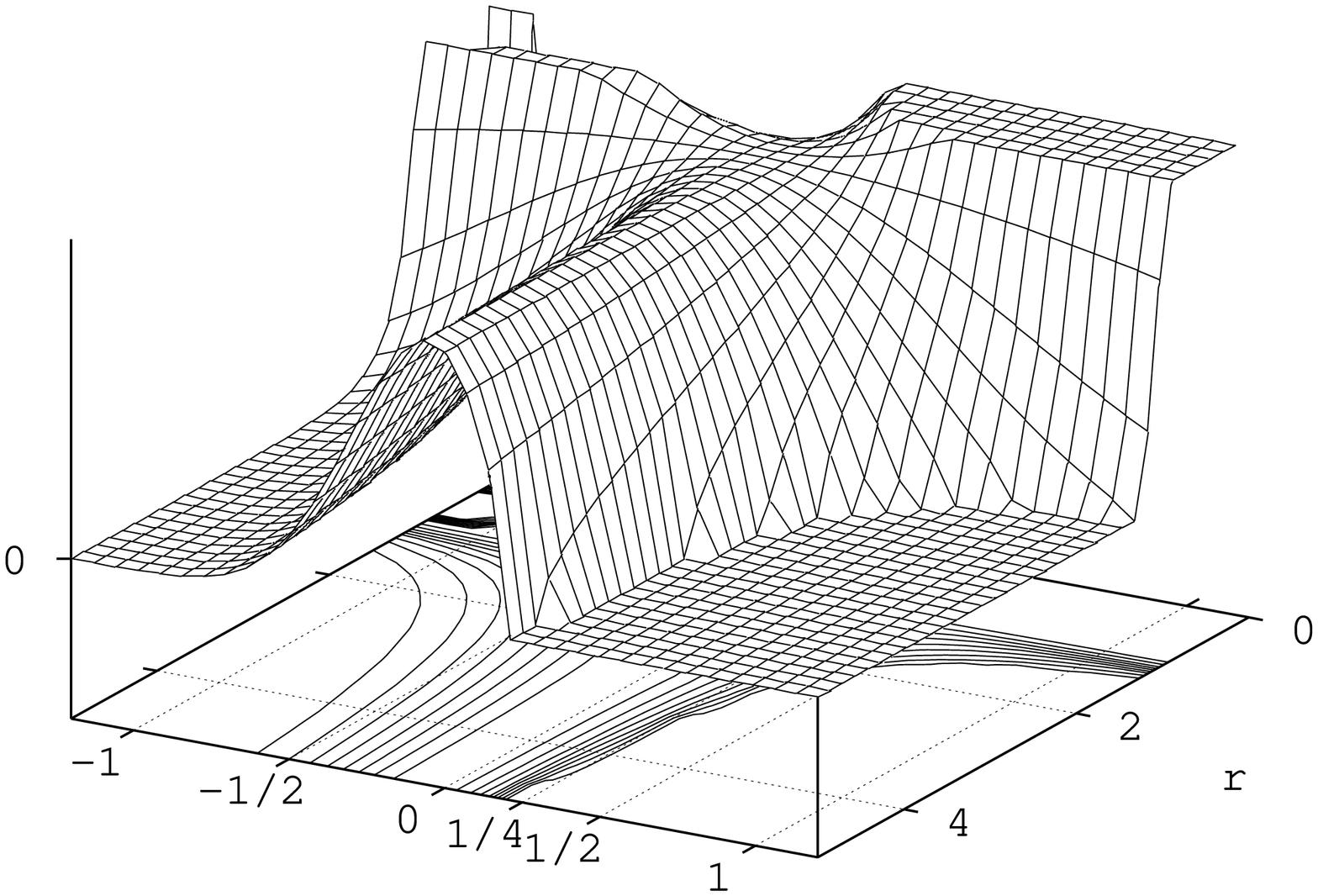}}}}

\centerline{(c)}
}
\hfill
\parbox[b]{7cm}{
\centerline{
\hbox{
   \epsfxsize=7cm
   \rotatebox{-90}{\epsfbox{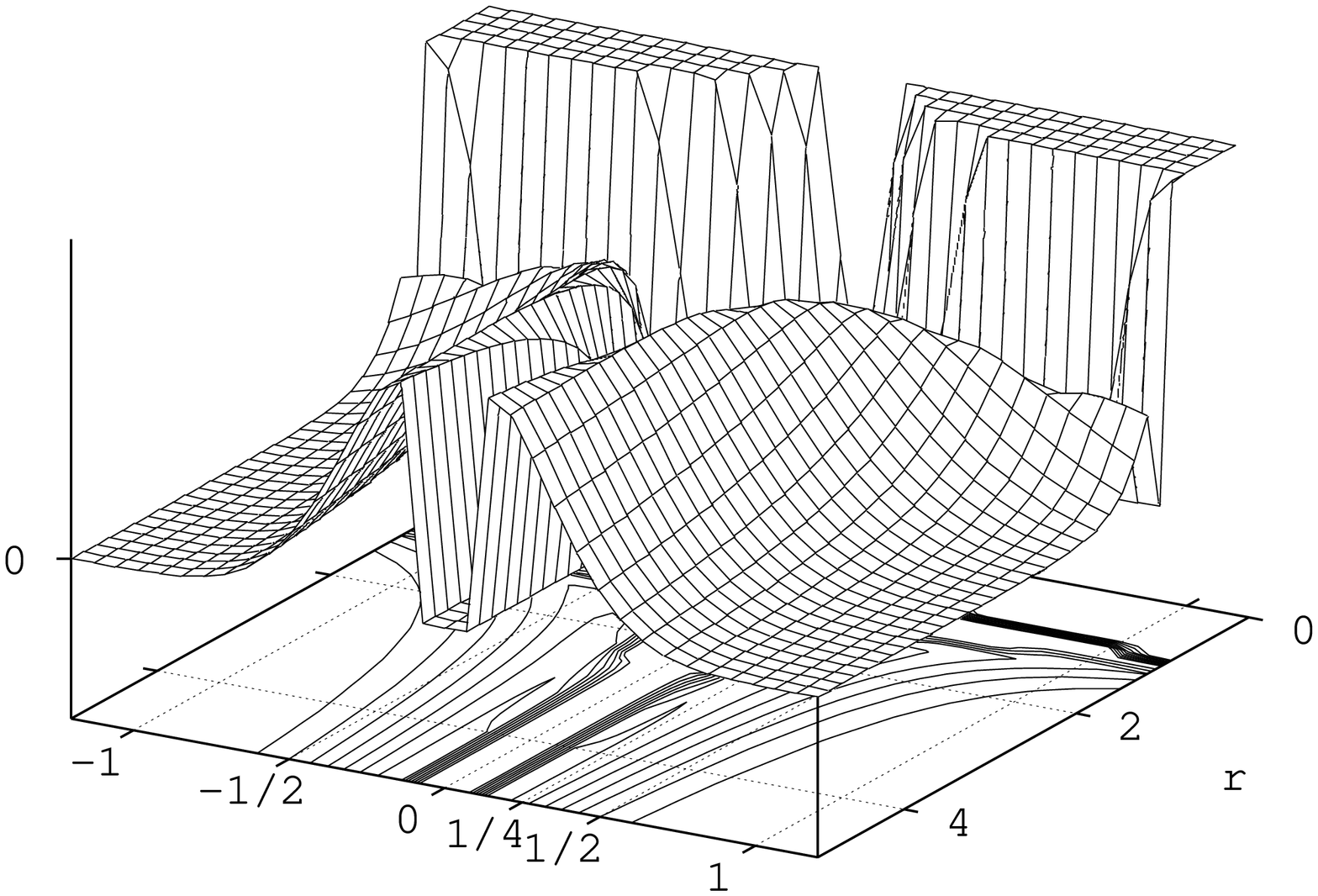}}}}

\centerline{(d)}
}
\caption{Absolute value of radial velocity for 
     the solution with radial electric field.
 (a) $u_{0}=1,\ q=0.1$.
 (b) $u_{0}=1,\ q=1$.
 (c) $u_{0}=2,\ q=0.1$.
 (d) $u_{0}=2,\ q=1$.	
 }
\label{fig-elw}
\end{figure}

The plots become slightly more complicated in presence of electric field in
Fig.~\ref{fig-elw}. The presence of another singularity in~(\ref{eq-dselw})
results in the fact that the radial geodesic motion of particles with
given energy is restricted to two separated regions perceptible in all
subplots (a)-(d). In case of negative $\sigma$ the singularity gets an
attractive character (the parts with increasing radial velocity at the back
of subplots). Moreover, in case of positive $\sigma$ and stronger electric
field (subplots (b),(d)) the radial motion is possible also at larger
radial distances from the $z$-axis (compare to corresponding subplots in
Figs.~\ref{fig-mgph},\ref{fig-mgz} which are different). In
subplots~\ref{fig-elw}(a),(c) we can again recognize motion with a constant
radial velocity for $\sigma=0$, when (\ref{eq-dselw}) becomes flat.

\begin{figure}[ht!]
\parbox[b]{7cm}{
\centerline{
\hbox{
   \epsfxsize=7cm
   \rotatebox{-90}{\epsfbox{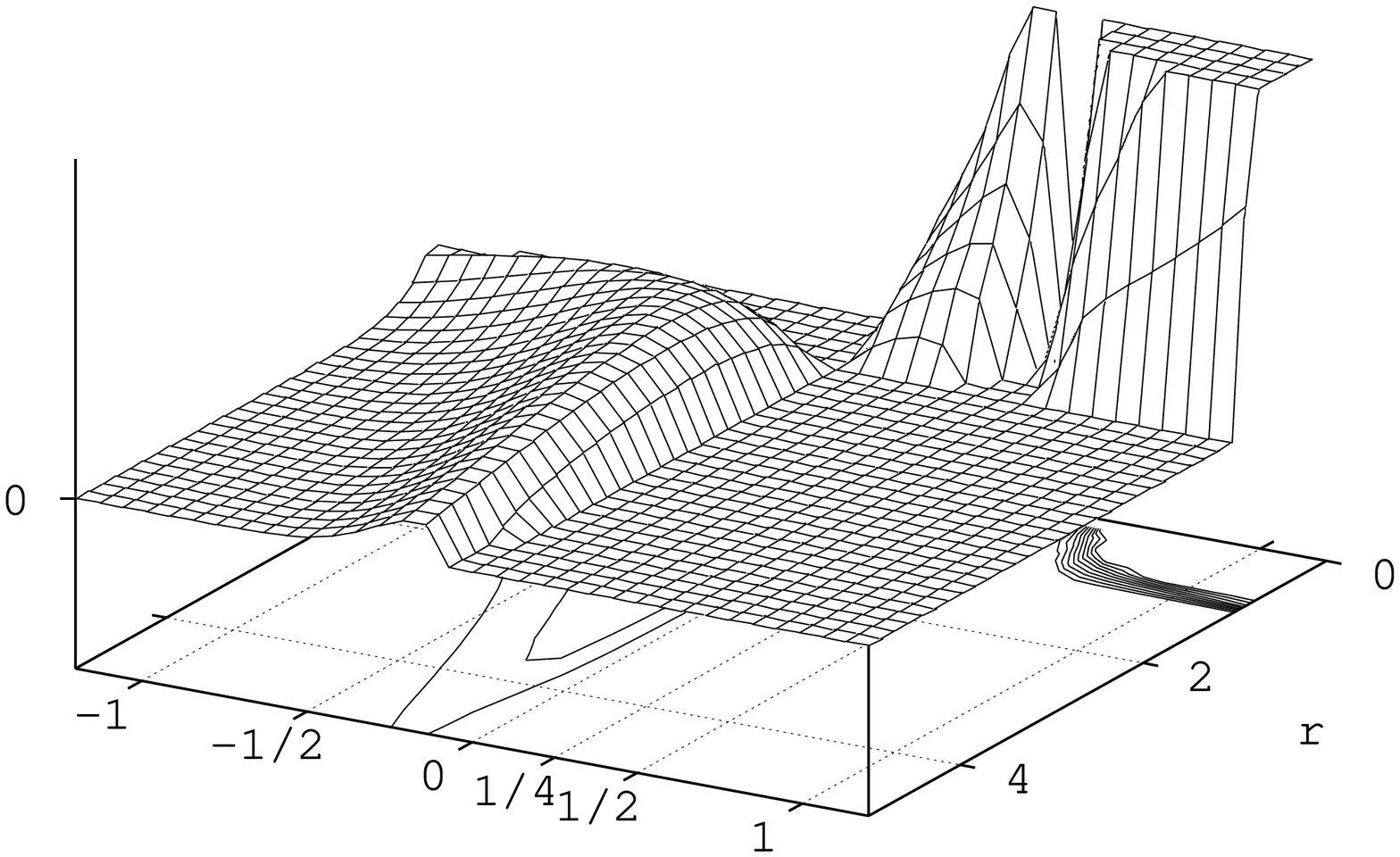}}}}

\centerline{(a)}
}
\hfill
\parbox[b]{7cm}{
\centerline{
\hbox{
   \epsfxsize=7cm
   \rotatebox{-90}{\epsfbox{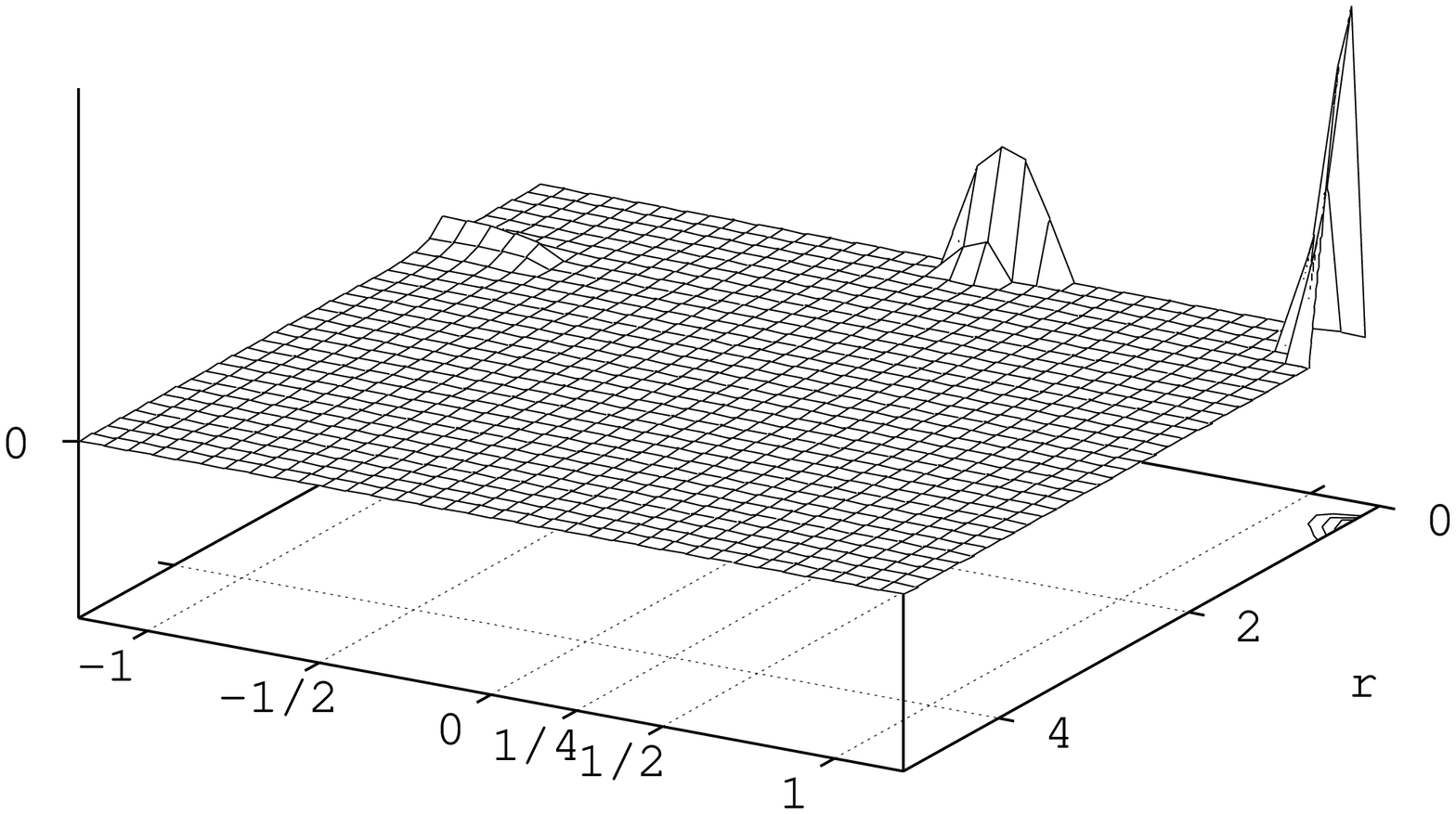}}}}

\centerline{(b)}
}

\vspace{5ex}
% -----------------------------------------------
\parbox[b]{7cm}{
\centerline{
\hbox{
   \epsfxsize=7cm
   \rotatebox{-90}{\epsfbox{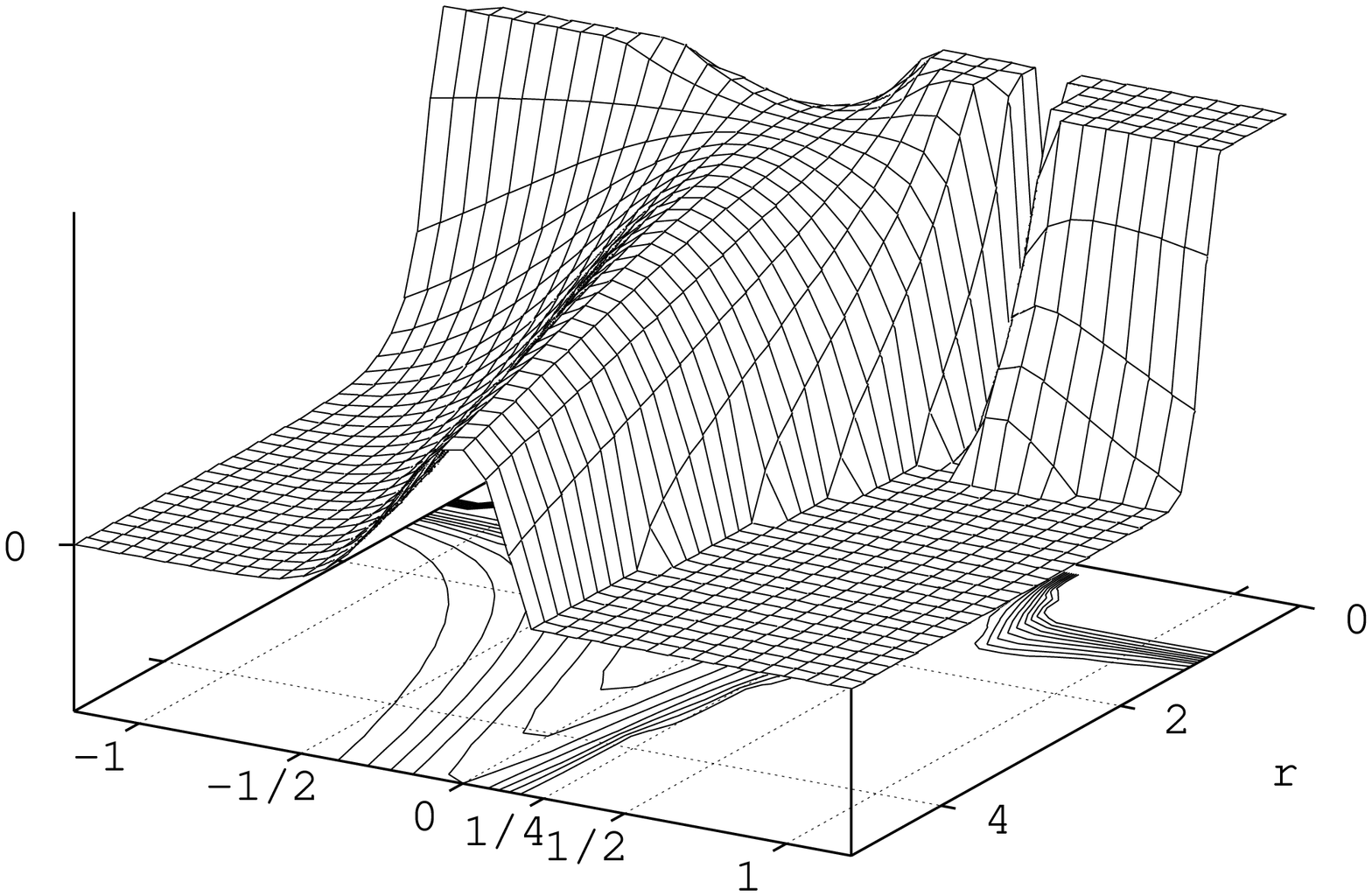}}}}

\centerline{(c)}
}
\hfill
\parbox[b]{7cm}{
\centerline{
\hbox{
   \epsfxsize=7cm
   \rotatebox{-90}{\epsfbox{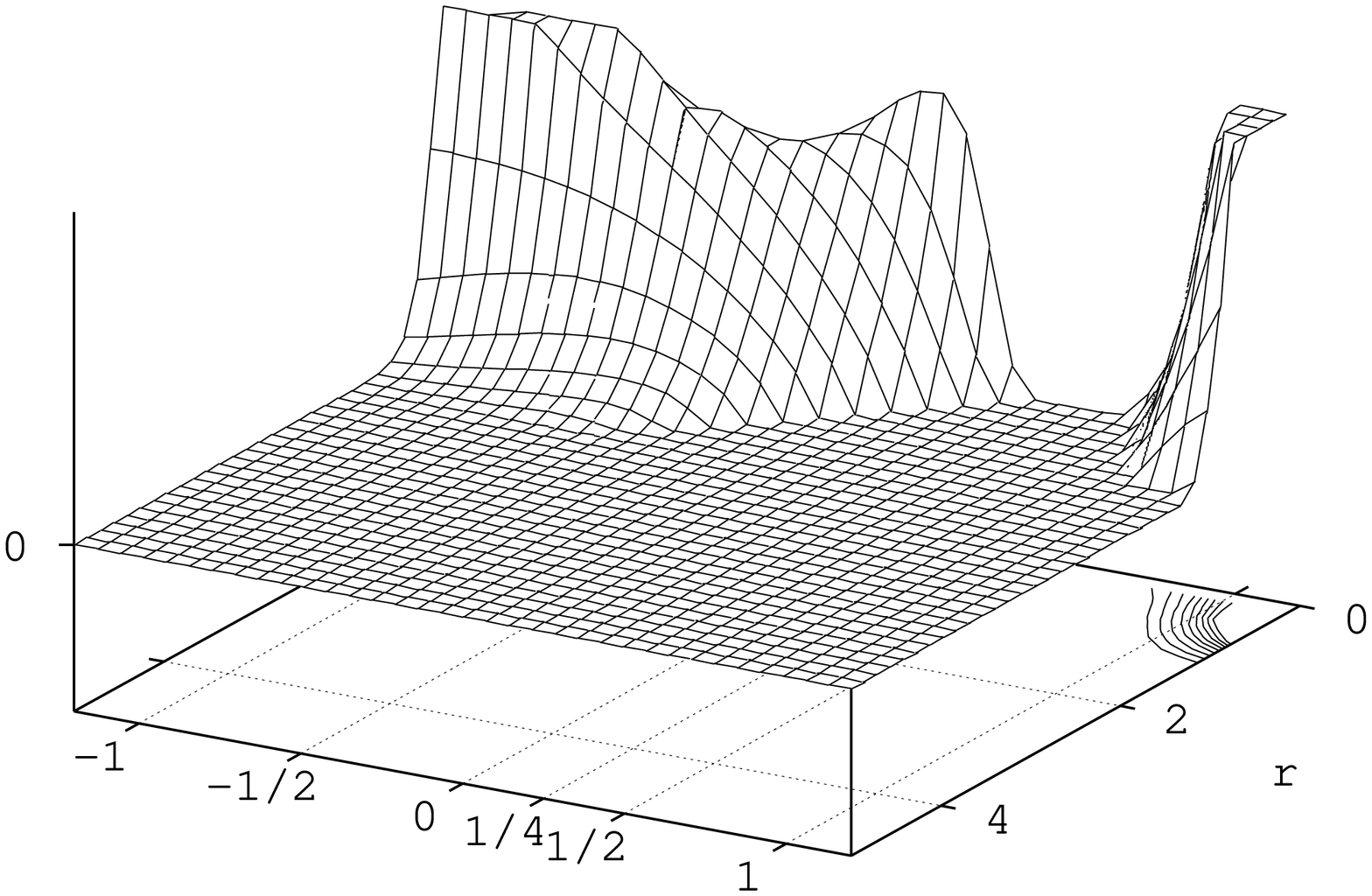}}}}

\centerline{(d)}
}
\caption{Absolute value of radial velocity for 
     the solution with longitudinal electric field.
 (a) $u_{0}=1,\ q=0.1$.
 (b) $u_{0}=1,\ q=1$.
 (c) $u_{0}=2,\ q=0.1$.
 (d) $u_{0}=2,\ q=1$.	
 }
\label{fig-elz}
\end{figure}

The solution including the longitudinal electric field
(section~\ref{sec-elwz}) has many features in common with the corresponding
magnetic one (\ref{eq-dsmgz}). Comparing Figs.~\ref{fig-mgz} and
\ref{fig-elz} one finds out that the electric case in Fig.~\ref{fig-elz}
differs in the particular detail that for the case of flat spacetime $\sigma=1/2$ the
radial velocity equals zero and the plane of constant $\sigma=1/2$ strictly
divides the surfaces on all subplots into two parts. The subplots
\ref{fig-elz}(a), (c) for a weak longitudinal electric field are analogous
to those ones corresponding to a weak longitudinal magnetic field in
Fig.~\ref{fig-mgz}(a), (c).

There should be stressed one more interesting point in connection with
Fig.~\ref{fig-elz}(b). For stronger electromagnetic field, even for positive
$\sigma$, the singularity qualitatively changes its behaviour: there is a
turning point close to the $z$-axis so that the particle cannot reach the
singularity (see the right part of the subplot (b)). This might be
another argument supporting the problematic interpretation of the LC
solution for $\sigma > 1$ (and consequently, all the solutions generated
from the LC metric in this paper).

Our discussion of the radial geodesic motion is, of course, far from
being exhaustive. Its aim is to summarize the physical interpretation of
the generated spacetimes and to emphasize the most essential points. The
physical qualities of those spacetimes are determined most of all by the
character of the seed metric. This conclusion is in full accordance with the
principles of the HM conjecture: all the generated EM fields represent a
generalization of the seed metric that must be their limiting case for a zero
electromagnetic field.

% -------------------- BIBLIOGRAPHY ---------------------------

% ------------------------------------------------------------------------
% F I G U R E S
% ------------------------------------------------------------------------
% -----------------------------------------------
% Figure 1
% -----------------------------------------------
% \begin{figure}[ht]
% \parbox[b]{6cm}{
% \centerline{
% \hbox{
%    \epsfxsize=6cm
%    \rotatebox{-90}{\epsfbox{lc1.eps}}}}

% \centerline{(a)}
% }
% \hfill
% \parbox[b]{6cm}{
% \centerline{
% \hbox{
%    \epsfxsize=6cm
%    \rotatebox{-90}{\epsfbox{lc2.eps}}}}

% \centerline{(b)}
% }
% \caption{Absolute value of radial velocity for the Levi-Civita solution.
%  (a) $u_{0}=1 $.
%  (b) $u_{0}=2 $.
% }
% \label{fig-lc}
% \end{figure}
% -----------------------------------------------
% Figure 2
% -----------------------------------------------
% -----------------------------------------------
% Figure 3
% -----------------------------------------------
% -----------------------------------------------
% Figure 4
% -----------------------------------------------
% -----------------------------------------------
% Figure 5
% -----------------------------------------------
% -----------------------------------------------
\end{document}